\documentclass[12pt]{article}
\newsavebox{\ns}
\newsavebox{\dbrane}
\newsavebox{\dbshort}

\usepackage{epsfig}
\usepackage{amssymb}

\def\appendix{{\newpage\section*{Appendix}}\let\appendix\section%
        {\setcounter{section}{0}
        \gdef\thesection{\Alph{section}}}\section}

\def\be{\begin{equation}}
\def\ee{\end{equation}}
\def\ba{\begin{eqnarray}}
\def\ea{\end{eqnarray}}

\newcommand{\nn}{\nonumber}

\newcommand{\ft}[2]{{\textstyle\frac{#1}{#2}}}
\newcommand{\eqn}[1]{(\ref{#1})}

\def\Dslash{\,\,{\raise.15ex\hbox{/}\mkern-12mu D}}
\def\Dbarslash{\,\,{\raise.15ex\hbox{/}\mkern-12mu {\bar D}}}
\def\delslash{\,\,{\raise.15ex\hbox{/}\mkern-9mu \partial}}
\def\delbarslash{\,\,{\raise.15ex\hbox{/}\mkern-9mu {\bar\partial}}}
\def\pslash{\,\,{\raise.15ex\hbox{/}\mkern-9mu p}}
\def\calDslash{\,\,{\raise.15ex\hbox{/}\mkern-12mu {\cal D}}}

\newcommand\mf{\mathcal{F}}
\newcommand\ma{\mathcal{A}}
\newcommand\mq{\mathcal{Q}}
\newcommand\mr{\mathbb{R}}
\newcommand\mcp{\bf \mathbb{C}P}
\newcommand\mhp{\bf \mathbb{H}P}
\newcommand\Sfour{\Sigma^{-}{\bf S}^4}
\newcommand\Sfive{\mathbb{R}^3 \times {\bf S}^5}
\newcommand\mh{\mathbb{H}}
\newcommand\ind{\mathrm{ind}}
\newcommand\Index{\mathrm{Index}}
\newcommand\myOverwrite[2]{\makebox[0cm][l]{#1}#2\ } 
\newcommand\D{\myOverwrite{D}{\slash}}

\begin{document}

\begin{titlepage}

\begin{center}
\today
{\small\hfill hep-th/0207244}\\
{\small\hfill MIT-CTP-3284}\\
{\small\hfill HUTP-02/A014}\\
{\small\hfill DAMTP-2002-85}\\

\vskip 1.5 cm
{\large \bf Conifold Transitions and Five-Brane Condensation in} \\
\vskip 0.2 cm
{\large \bf M-Theory on Spin(7) Manifolds} \\

\vskip 1.5 cm
{Sergei Gukov$^1$, James Sparks$^2$, and David Tong$^3$}\\
\vskip 1cm

$^1${\sl Jefferson Physical Laboratory, Harvard University, \\
Cambridge, MA 02138, U.S.A. \\ {\tt gukov@democritus.harvard.edu}\\}
\vskip 0.5cm
$^2${\sl Centre for Mathematical Sciences, University of Cambridge, \\
Wilberforce Road, Cambridge CB3 0WA, UK. \\
{\tt J.F.Sparks@damtp.cam.ac.uk}\\}
\vskip 0.5cm
$^3${\sl Center for Theoretical Physics,
Massachusetts Institute of Technology, \\ Cambridge, MA 02139, U.S.A.
\\ {\tt dtong@mit.edu}\\}

\end{center}

\vskip 0.5 cm
\begin{abstract}
We conjecture a topology changing transition in M-theory
on a non-compact asymptotically conical $Spin(7)$ manifold,
where a 5-sphere collapses and a ${\bf \mathbb{C}P}^2$ bolt grows. 
We argue that the transition may be understood as the condensation of
M5-branes wrapping ${\bf S}^5$. Upon reduction to ten dimensions,
it has a physical interpretation as a transition of D6-branes lying 
on calibrated submanifolds of flat space. In yet another guise, it 
may be seen as a geometric transition 
between two phases of type IIA string theory on a $G_2$ holonomy 
manifold with either wrapped D6-branes, or background Ramond-Ramond flux.
This is the first non-trivial example of a topology changing transition
with only $1/16$ supersymmetry.

\end{abstract}

\end{titlepage}

\pagestyle{plain}
\setcounter{page}{1}
\newcounter{bean}
\baselineskip16pt
\tableofcontents


\section{Introduction}

Topology changing transitions in string theory are of great interest
\cite{brian}. These have been well studied in compactifications of
type II string theory on Calabi-Yau manifolds, where the residual
${\cal N}=2$ supersymmetry provides much control over the dynamics.
There are two prototypical examples. The flop transition, in
which a two-cycle shrinks and is replaced by different two-cycle,
proceeds smoothly in string theory \cite{greene,witten}.
In contrast, the conifold transition, in which a three-cycle shrinks
and a two-cycle emerges, is accompanied by a phase transition in
the low-energy dynamics which can be understood as the condensation
of massless black holes \cite{Strominger,gms}.

In the past year, there has been great progress in understanding similar
effects in compactifications of M-theory on manifolds of $G_2$ holonomy,
where the resulting four dimensional theories have ${\cal N}=1$
supersymmetry. There is, once again, an analog of the flop transition;
this time three-cycles shrink and grow and,
as with the Calabi-Yau example, the process is smooth \cite{amv,aw}.
Other $G_2$ geometrical transitions involving shrinking $\mcp^2$'s have
also been discussed \cite{aw}. These proceed via a phase
transition but, unlike the conifold transition, do not appear to
be related to condensation of any particle state\footnote{However,
we shall argue below that this interpretation can be given to the
same transition in type IIA string theory.}. For related work, see 
\cite{enmass}. 

\begin{figure}
\setlength{\unitlength}{0.9em}
\begin{center}
\begin{picture}(25,10)

\put(10,2){\line(1,0){6}}
\put(16,2){\line(1,1){2}}
\put(10,2){\line(1,2){3}}
\put(16,2){\line(-1,2){3}}
\put(18,4){\line(-5,4){5}}
\put(13,0.8){${\bf S}^3$}
\put(17,2){${\bf S}^2$}
\put(11,-1){$m_{q}=0$}
\put(12,10){conifold}
\put(12,9){singularity}

\put(6,6){$\Longleftarrow$}
\put(5,7){deformation}

\put(-2,2){\line(1,0){6}}
\put(4,2){\line(1,1){2}}
\put(-2,2){\line(1,3){2}}
\put(4,2){\line(-1,3){2}}
\put(0,8){\line(1,0){2}}
\put(6,4){\line(-1,1){4}}
\put(1,0.8){${\bf S}^3$}
\put(5,2){${\bf S}^2$}
\put(-1,-1){$m_{q} \ne 0$}
\put(1,9){${\bf S}^3$}

\put(21,6){$\Longrightarrow$}
\put(19,7){resolution}

\put(22,2){\line(1,0){6}}
\put(28,2){\line(1,1){2}}
\put(22,2){\line(3,5){3}}
\put(28,2){\line(-3,5){3}}
\put(25,7){\line(1,1){2}}
\put(30,4){\line(-3,5){3}}
\put(25,0.8){${\bf S}^3$}
\put(29,2){${\bf S}^2$}
\put(23,-1){$\langle q \rangle \ne 0$}
\put(24.5,8.5){${\bf S}^2$}

\end{picture}\end{center}
\caption{Conifold transition in type IIB string theory.}
\label{figa}
\end{figure}
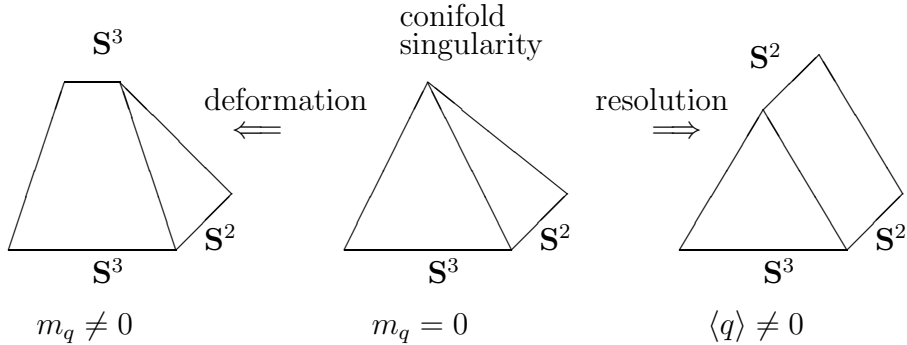

The purpose of this paper is to study geometrical transitions 
in M-theory on eight-dimensional manifolds 
with $Spin(7)$ holonomy.
Since the physics is very similar to the conifold
transition in Calabi-Yau manifolds,
let us briefly recall what happens in that case.
As the name indicates, the conifold is a cone over a five dimensional space
which has topology ${\bf S}^2 \times {\bf S}^3$ (see Figure \ref{figa}).
Two different ways to desingularize this space --- called
the deformation and the resolution --- correspond to replacing the 
singularity by a finite size ${\bf S}^3$ or ${\bf S}^2$, respectively.
In type IIB string theory, the two phases of the 
conifold geometry correspond to different
branches in the four-dimensional 
$\mathcal{N}=2$ low-energy effective field theory.
In the deformed conifold phase, D3-branes wrapped around the 
3-sphere give rise to a low-energy field $q$, with mass determined 
by the size of the ${\bf S}^3$. 
In the effective four-dimensional supergravity theory these
states appear as heavy, point-like, extremal black holes.
On the other hand, in the resolved conifold phase the field $q$
acquires an expectation value reflecting the condensation of 
these black holes. 
Of course, in order to make the transition from one phase
to the other, the field $q$ must become massless somewhere and this 
happens at the conifold singularity, as illustrated 
in Figure \ref{figa}.

In this paper we will argue that a similar phenomenon occurs in
M-theory on a $Spin(7)$ manifold with a certain conical singularity.
Apart from related orbifold constructions, there are essentially 
only two types
of conical singularity which are known, at present, to admit
a resolution to a smooth complete metric with $Spin(7)$ holonomy. These are listed in 
Table 1.
The first corresponds to a cone over $SO(5)/SO(3) = {\bf S}^7$
and was constructed a long time ago in \cite{gary}.
The resolution of this conical singularity leads to
a smooth non-compact $Spin(7)$ manifold isomorphic to
an $\mathbb{R}^4$ bundle over ${\bf S}^4$. Extending the 
ansatz to asymptotically locally conical (ALC) metrics, in 
which a circle stabilizes at finite size asymptotically, it 
was shown that \cite{gary7} there are two families of topologically distinct 
resolutions of this cone, labelled 
$\mathbb{B}_8\cong\mr^4\times{\bf S}^4$ and 
$\mathbb{A}_8 \cong \mathbb{R}^8$. One might therefore expect 
that $\mathbb{A}_8$ and $\mathbb{B}_8$ are different phases of M-theory
on the same conical singularity. However, we shall argue below that 
this is {\it not} the case.

\begin{table}\begin{center}
\begin{tabular}{|c|c|c|}
\hline
$X$ & Topology & Base of Cone\\
\hline
\hline
$\mathbb{R}^4 \times {\bf S}^4$ & chiral spin bundle of 
${\bf S}^4$ & ${\bf S}^7 =
SO(5)/SO(3)$ \\
\cline{1-3}
$\mathbb{R}^4 \times {\bf \mathbb{C}P}^2$ & universal quotient bundle & $N_{1,-1}=SU(3)/U(1)$ \\
\cline{1-2}
$\mathbb{R}^3\times{\bf S}^5$ & $\mathbb{R}^3$ bundle over ${\bf S}^5$
& \\
\hline
\end{tabular}\end{center}
\label{table}
\caption{The two cases of Spin(7) conical singularity studied in this paper.}
\end{table}

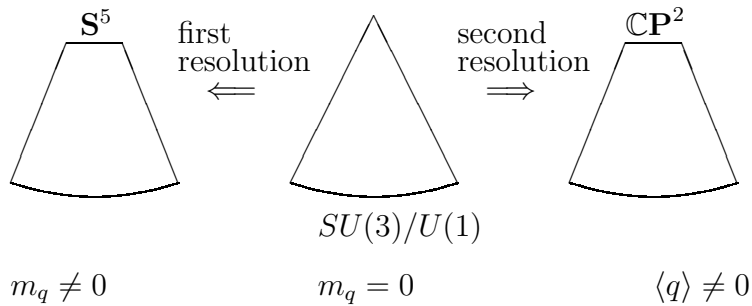
\begin{figure}
\setlength{\unitlength}{0.9em}
\begin{center}
\begin{picture}(25,11)

\qbezier(10,3)(13,2)(16,3)
\put(10,3){\line(1,2){3}}
\put(16,3){\line(-1,2){3}}
\put(11,1.1){$SU(3)/U(1)$}
\put(11,-1){$m_{q}=0$}

\qbezier(0,3)(3,2)(6,3)
\put(0,3){\line(2,5){2}}
\put(6,3){\line(-2,5){2}}
\put(2,8){\line(1,0){2}}
\put(0,-1){$m_{q} \ne 0$}
\put(2.5,8.3){${\bf S}^5$}

\qbezier(20,3)(23,2)(26,3)
\put(20,3){\line(2,5){2}}
\put(26,3){\line(-2,5){2}}
\put(22,8){\line(1,0){2}}
\put(23,-1){$\langle q \rangle \ne 0$}
\put(22,8.3){${\bf \mathbb{C}P}^2$}

\put(7,6){$\Longleftarrow$}
\put(6,8){first}
\put(6,7){resolution}
\put(17,6){$\Longrightarrow$}
\put(16,8){second}
\put(16,7){resolution}

\end{picture}\end{center}
\caption{Conifold transition in M-theory on a manifold
with $Spin(7)$ holonomy.}
\label{figb}
\end{figure}

The second conical singularity discussed in the literature
corresponds to a cone over $SU(3)/U(1)$ \cite{GS,garycohom,kanno}.
It is in this case that we suggest an interesting phase transition.
We conjecture that there exist two possible ways of resolving this singularity,
illustrated in Figure \ref{figb}.
A well-known resolution consists of gluing in a copy
of ${\bf \mathbb{C}P}^2$ in place of the singularity.
This leads to a one-parameter family of complete
metrics with $Spin(7)$ holonomy on the universal
quotient bundle $\mathcal{Q}$ of ${\bf \mathbb{C}P}^2$, labelled 
by the volume of the ${\bf \mathbb{C}P}^2$ bolt. They have 
topology,
\be
\mathcal{Q} \cong \mathbb{R}^4 \times {\bf \mathbb{C}P}^2
\label{qtop}
\ee
A less well-known resolution of this $Spin(7)$ conifold
may be obtained by blowing up a copy of the five-sphere.
Some numerical evidence for the existence of such a metric 
was presented in \cite{garycohom}. It remains an open 
problem to find an explicit $Spin(7)$ metric 
with these properties; a way to approach this, and a review of the 
known results, is presented in the appendix.
In Section 2, using the relationship 
with singularities of coassociative submanifolds in $\mathbb{R}^7$,
we provide further strong evidence for the existence of a complete
$Spin(7)$ metric with an ${\bf S}^5$ bolt:
\be
X\cong\Sfive
\label{sfivetop}
\ee
Furthermore, we hypothesize  that the manifolds with topology 
\eqn{qtop} and \eqn{sfivetop} are analogous to the resolution 
and deformation of the conifold, respectively. 
In other words, (\ref{qtop}) and (\ref{sfivetop}) are two
phases of what one might call {\it a $Spin(7)$ conifold}.

As with the conifold transition \cite{Strominger,gms}, the topology 
changing transition in M-theory on the $Spin(7)$ cone over 
$SU(3)/U(1)$ has a nice interpretation in terms of the 
low-energy effective field theory. We argue that the effective 
dynamics of M-theory on the cone over $SU(3)/U(1)$ is described by a
three-dimensional ${\cal N}=1$ abelian Chern-Simons-Higgs theory. 
The Higgs field $q$ arises upon quantization of the M5-brane 
wrapped over the ${\bf S}^5$. We 
propose that, at the conifold point where the five-sphere shrinks, 
these M5-branes become massless as suggested by the classical geometry. 
At this point, the theory may pass through a phase transition into 
the Higgs phase, associated with the condensation of these five-brane 
states. 

To continue the analogy with the Calabi-Yau conifold, recall that the 
moduli space of type II string theory on the Calabi-Yau 
conifold has three semi-classical regimes. The deformed  
conifold provides one of these, while there are two 
large-volume limits of the resolved conifold, related 
to each other by a flop transition. In fact, the same picture 
emerges for 
the $Spin(7)$ conifold. In this case, however, the two backgrounds 
differ not in geometry, but in the G-flux. It was shown in 
\cite{GS} that, due to the membrane anomaly of \cite{witten1}, 
M-theory on $\mathcal{Q}\cong\mathbb{R}^4 \times{\bf \mathbb{C}P}^2$ 
is consistent only for half-integral units of G$_4$ through 
the $\mcp^2$ bolt. We will show that, after the transition 
from $X\cong\Sfive$, the G-flux may take the values $\pm 1/2$, 
with the two possibilities related by a parity transformation. Thus, 
the moduli space of M-theory on the $Spin(7)$ cone over $SU(3)/U(1)$ 
also has three semi-classical limits: one with the parity invariant 
background geometry $\Sfive$, and two with the background geometry 
$\mathcal{Q}$ which are mapped into each other under parity.

In view of the interesting phenomena associated to branes in the 
conifold geometry, and their relationship to the conifold transition 
\cite{klebwitt,KStrassler}, it would be interesting 
to learn more about the $Spin(7)$ transition using membrane probes in 
this background, and also to study the 
corresponding holographic renormalization group flows. For work in this 
area, see \cite{gt,oz,carlos}.

Finally, we would like to mention a second 
interpretation of the $Spin(7)$ topology  
changing transition, which again has an analog among 
lower dimensional manifolds. To see this, let us first 
recall the story of the $G_2$ flop \cite{amv,aw} and its 
relationship to the brane/flux duality of the conifold \cite{vafa}. 
In this scenario, one starts in type IIA theory with 
the familiar geometry of the deformed conifold, and 
wraps an extra D6-brane around the 3-cycle. This yields 
a system with $\mathcal{N}=1$ supersymmetry in 3+1 dimensions. 
A natural question one could ask is: ``What happens if one 
tries to go through the conifold transition with the 
extra D6-brane?''. One possibility could be that the other 
branch is no longer connected and the transition is not possible.
However, this is not what happens. Instead the physics is somewhat 
more interesting. According to \cite{vafa,amv,aw},
the transition proceeds, but now the two branches are smoothly 
connected, with the wrapped D6-brane replaced by RR 2-form 
flux through the ${\bf S}^2$. Since both  D6-branes and 
RR 2-form tensor fields lift to purely geometric backgrounds in 
M-theory, the geometric transition can be understood as
a flop-like transition in M-theory on a $G_2$ manifold:
$$
X \cong \mr^4 \times {\bf S}^3
$$
For example, to obtain the resolved conifold with RR 2-form flux
one can choose the `M-theory circle' to be the fiber of
the Hopf bundle (see \cite{kmpt} for a recent discussion)
$$
{\bf S}^1 \hookrightarrow {\bf S}^3 \to {\bf S}^2
$$
while choosing instead an embedding of the M-theory circle in 
$\mr^4$ gives rise the deformed conifold, with the D6-brane 
localized on the ${\bf S}^3$ fixed point set \cite{bobby,amv}.

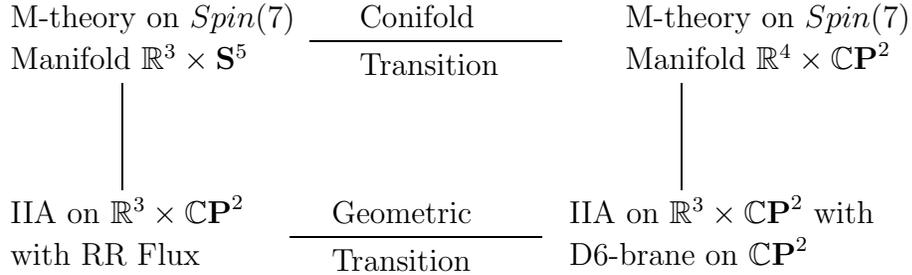
\begin{figure}
\setlength{\unitlength}{0.9em}
\begin{center}
\begin{picture}(25,10)
\put(0,9){M-theory on $Spin(7)$}\put(0,7.5){Manifold $\Sfive$}
\put(22,9){M-theory on $Spin(7)$}
\put(22,7.5){Manifold $\mathbb{R}^4 \times {\bf \mathbb{C}P}^2$}
\put(0,2){IIA on $\mathbb{R}^3 \times {\bf \mathbb{C}P}^2$}
\put(0,0.5){with RR Flux}
\put(20,2){IIA on $\mathbb{R}^3 \times {\bf \mathbb{C}P}^2$ with}
\put(20,0.5){D6-brane on ${\bf \mathbb{C}P}^2$}
\put(10.7,8.5){\line(1,0){9}}\put(12.5,9){Conifold}\put(12.5,7.3){Transition}
\put(4,7){\line(0,-1){3.8}}
\put(24,7){\line(0,-1){3.8}}
\put(10,1.5){\line(1,0){9}}\put(11.5,2){Geometric}\put(11.5,0.3){Transition}
\end{picture}\end{center}
\caption{Geometric transition as a conifold transition in
M-theory on $Spin(7)$ manifold.}
\label{figc}
\end{figure}

As we now explain, this 
procedure works in a slightly different and interesting way for 
our $Spin(7)$ manifolds with less supersymmetry. Our starting 
point is type IIA string theory on the $G_2$ holonomy manifold
$$
M^7 \cong \mathbb{R}^3 \times {\bf \mathbb{C}P}^2
$$
which is obtained by resolving the cone over $SU(3)/U(1)^2$.
As we shall see in Section 3, the effective low-energy theory is 
an $\mathcal{N}=2$ supersymmetric abelian Higgs model 
in 2+1 dimensions, and its dynamics is very similar to 
compactification of M-theory on the same manifold $M^7$ \cite{aw}. 
In particular, the quantum moduli space consists of three branches, 
each of which arises from compactification on a manifold of topology $M^7$, 
connected by a singular phase transition. Following the ideas of 
\cite{vafa,amv}, one could wrap an extra D6-brane over the $\mcp^2$ and ask
a similar question: ``What happens if one tries
to go through a phase transition?''. 
Using arguments similar to \cite{amv,Gomis}, we conjecture that the 
transition is again possible, via M-theory on a $Spin(7)$ manifold.
More precisely, we claim that after the geometric transition
one finds type IIA string theory on $M^7$, where
the D6-brane is replaced by RR flux through $\mcp^1\subset\mcp^2$.
This leads to a fibration:
$$
{\bf S}^1 \hookrightarrow {\bf S}^5 \to \mcp^2
$$
Hence the M-theory lift of this configuration gives
a $Spin(7)$ manifold with the topology $\Sfive$.
Similarly, one can identify the lift of $M^7$ with
a D6-brane wrapped around $\mcp^2$ as the $Spin(7)$ manifold
$\mathcal{Q} \cong \mathbb{R}^4 \times {\bf \mathbb{C}P}^2$.
Summarizing, we find that the conifold transition in M-theory
on a $Spin(7)$ manifold is nothing but a geometric transition
in IIA string theory on the $G_2$ manifold $M^7$ with branes/fluxes,
as shown in Figure \ref{figc}. However, unlike the Calabi-Yau 
$\rightarrow G_2$ 
example, in our case of $G_2\rightarrow Spin(7)$, the transition 
does not proceed smoothly in M-theory.

The paper is organized as follows. In the following section, 
we study the relationship between D6-branes on coassociative 
submanifolds of $\mr^7$, and their lift to M-theory on manifolds 
of $Spin(7)$ holonomy. We demonstrate the conifold transition 
explicitly from the D6-brane perspective. We further discuss 
several aspects of M-theory on $Spin(7)$ manifolds, including 
fluxes, anomalies, parity and supersymmetry.  In Section 3 we 
turn to the interpretation of the conifold transition from 
the low-energy effective action. We build a consistent picture 
in which the geometric transition is understood as a Coulomb to Higgs 
phase transition in a Chern-Simons-Higgs model. Finally, in Section 4, 
we discuss further aspects of the geometry, and the different reductions 
to type IIA string theory by quotienting the $Spin(7)$ manifolds. 
We include explicit constructions of the relevant quotient for the 
brane/flux transition, as well as the D6-brane loci of Section 2. 
In the appendix, we review the current state of knowledge for the geometry 
of $X\cong \Sfive$, and use the methods of Hitchin \cite{Hitchin} 
to determine properties of the metric.


\section{D6-Branes, M-Theory and $Spin(7)$ Conifolds}

Our main interest in this section is to understand the geometric
transition between the asymptotically conical (AC) $Spin(7)$ manifolds
$\mq$ and $\Sfive$, both of which are resolutions of the cone on the
weak $G_2$ holonomy Aloff-Wallach space, $N_{1,-1}=SU(3)/U(1)$. 
Details of the metrics on these spaces are reviewed in the appendix.
Here we start by understanding the 
transition through the relationship to D6-branes spanning 
coassociative submanifolds of $\mathbb{R}^7$. 

The key observation is that D6-branes lift to pure geometry in M-theory 
\cite{paul}. We start with a configuration of D6-branes in flat 
Minkowski space $\mr^{1,9}$, with worldvolume $\mr^{1,2}\times L$. 
The branes preserve at least two supercharges (${\cal N}=1$ supersymmetry 
in $2+1$ dimensions) if 
we choose the four-dimensional locus $L\subset \mr^7$ to be a 
coassociative submanifold \cite{becker}, calibrated by 
\be
\Psi^{(4)}=\star\Psi^{(3)}=e^{2457}+e^{2367}+e^{3456}+e^{1256}
+e^{1476}+e^{1357}+e^{1234}
\label{Psi}\ee
Upon lifting to M-theory, the D6-brane configuration becomes the background 
geometry $\mr^{1,2}\times X$ where $X$ is an eight-dimensional manifold 
equipped with a metric of $Spin(7)$ holonomy. When $L$ is smooth, matching 
of states in the IIA and M-theory descriptions leads to the homology 
relations between $L$ and $X$ \cite{GS},
\be
h_0(L)=h_2(X)+1,\ \ \ \ \ H_i(L,\mathbb{Z})\cong H_{i+2}(X,\mathbb{Z})\ \ 
\ \ \  \  \ \ i>0
\label{homo}\ee
These relations imply, among other things, that the Euler numbers
of $X$ and $L$ should be the same. In all our examples one can
easily check that this is indeed true, say, via cutting out $L$
inside $X$ and showing that the Euler number of the remaining
manifold with boundary is zero.

In this section we examine the coassociative four-fold geometry discussed 
by Harvey and Lawson \cite{HL}. Using the equations \eqn{homo}, we show 
that certain transitions between D6-branes can be reinterpreted as 
geometric transitions between $Spin(7)$ manifolds of the type described 
in the introduction.
Of course, one can also work in reverse and, given M-theory on a non-compact 
$Spin(7)$ manifold $X$, we may attempt to find a IIA description in terms of 
D6-branes on coassociative 
submanifolds of $\mr^7$. This is possible if $X$ admits a $U(1)$ isometry, 
which we identify as the M-theory circle, such that the quotient becomes,
\be
X/U(1)\cong \mathbb{R}^7
\label{quotient}\ee
In this case, all information about the topology of $X$ is stored in
the fixed point set $L$. This set has an interpretation 
as the locus of D6-branes in type IIA. The task of identifying $L$
given $X$ is somewhat involved and we postpone the calculations to Section 4,
where we explicitly construct the quotient to find the locus $L$.

\subsection{D6-Branes and the Cone over $SO(5)/SO(3)$}

Before we examine the example relevant for the conifold transition, 
let us first start with the $Spin(7)$ holonomy metric on the cone over 
${\bf S}^7\cong SO(5)/SO(3)$. As we shall see, 
the coassociative locus $L$ for this example is intimately related to the 
$N_{1,-1}$ case of primary interest.  
The $Spin(7)$ holonomy metric on the cone over ${\bf S}^7$ has a 
resolution to the chiral spin bundle of ${\bf S}^4$, 
\be 
X=\Sigma^-{\bf S}^4\cong\mr^4\times{\bf S}^4 
\label{myfriend}\ee
This manifold has isometry group $Sp(2)\times Sp(1)$. Since  $H^2(X;U(1))$ 
is trivial, M-theory compactified on 
$X$ has no further symmetries arising from the $C$-field \cite{GS}.
As we shall describe in detail in Section 4, the D6-brane locus $L$  
arises if we choose the M-theory circle to be embedded diagonally within 
the full isometry group. Upon reduction to type IIA string theory, 
this symmetry group is broken\footnote{The global structure is $U(2)$.}
to $Sp(1)\times U(1)$, which acts on the locus $L$.
The $Sp(1)$-invariant coassociative cones in $\mr^7$ can be
completely classified \cite{Ejiri,Malcev,Mashimo}. Without going
into details of these methods, we simply mention that the result 
derives from
classification of three-dimensional simple subalgebras in the Cayley algebra.
This leads, essentially, to two distinct families of coassociative
cones: one discussed by Harvey and Lawson \cite{HL},
and one constructed by Mashimo \cite{Mashimo}. As we will argue below,
it is the first case which is relevant to our problem.
We leave the analysis of the second case to the interested reader.

In order to describe the locus $L$, it will prove 
useful to decompose $\mr^7$ in terms of the quaternions $\mathbb{H}$, 
\be
\mathbb{R}^7 = \mathbb{R}^3 \oplus \mathbb{R}^4 = \mathrm{Im}
\mathbb{H} \oplus \mathbb{H}
\ee
The advantage of this notation is that it makes manifest a natural $Sp(1)$ 
action. To see this, let $x\in\mathrm{Im}\mathbb{H}$ and $y\in\mathbb{H}$. 
Then for $q\in Sp(1)$, we have the action
\be
q:(x,y)\rightarrow (qx\bar{q}, y \bar{q})\label{HLaction}
\ee
This acts on the $\mr^3$ factor as the usual $Sp(1)\sim SO(3)$ action. 
The action on the second factor $\mathbb{H}$ may be understood in terms 
of the usual action of $Spin(4)\sim SO(4)$ on
$\mathbb{R}^4$, where we write $Spin(4)=Sp(1)_L\times
Sp(1)_R$ and take $Sp(1)=Sp(1)_R$.


\begin{figure}
\begin{center}
\epsfxsize=5in\leavevmode\epsfbox{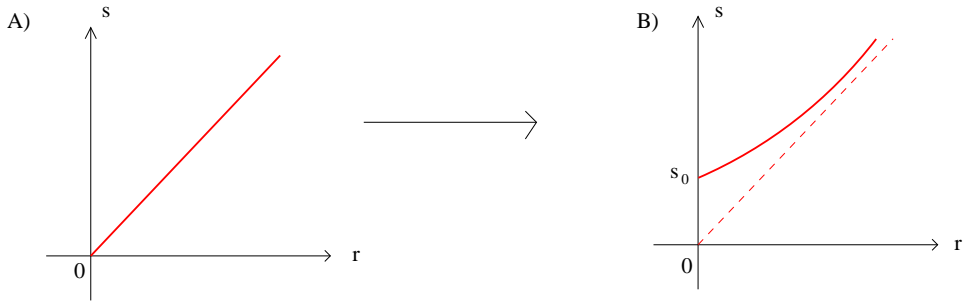}
\end{center}
\caption{A curve in the $r-s$ plane, whose $Sp(1)$ orbits sweep 
out coassociative submanifolds: A) the singular cone over a squashed 
three-sphere, and B) $L=H^1$.}
\label{hlfiga}
\end{figure}

The utility of the action \eqn{HLaction} on $\mathbb{R}^7=\mathrm{Im}\mathbb{O}$
lies in the fact that it preserves the $G_2$ structure. This fact was employed by 
Harvey and Lawson to construct $Sp(1)$-invariant calibrated submanifolds. 
Define a radial coordinate $s$ on $\mathrm{Im}\mathbb{H}$ and a radial 
coordinate $r$ 
on $\mathbb{H}$, and consider the curve in the  $r-s$ plane given by
\be
s(4s^2 - 5r^2)^2 = \label{curve}\rho\ge 0
\ee
Then it can be shown that, under the action of $Sp(1)$, we sweep out a 
coassociative submanifold $L$ of $\mathbb{R}^7$ \cite{HL}. 
To describe $L$ explicitly, let us introduce of a fixed unit 
vector $\epsilon\in \mathrm{Im}\mathbb{H}$. Then
\be
L = \{ (s q \epsilon \bar q, r \bar q)  ~:~~ q \in Sp(1),
\quad (s,r) \in \mathbb{R}^+ \times \mathbb{R}^+,
\quad s(4s^2 - 5r^2)^2 = \rho \}
\label{hungrynow}\ee
When the deformation parameter vanishes, $\rho=0$, the curve \eqn{curve} 
has two solutions. For now we will not consider the simplest branch, 
$s=0$, but instead restrict attention to
\be
s=  {\sqrt{5} \over 2} r
\label{lime}\ee
for which the coassociative four-fold $L$ described in \eqn{hungrynow} 
is a cone over the squashed three-sphere. It is on this submanifold 
that we place a single D6-brane, as depicted in Figure (\ref{hlfiga}A). 
Now consider resolving the conical singularity of $L$ by turning on 
$\rho>0$. Again, there are two branches, and we restrict attention 
to $s > \sqrt{5}r/2$ as shown in Figure (\ref{hlfiga}B). 
At large distances, $s,r\gg \rho$, $L$ is 
asymptotic to a cone over a squashed three-sphere. However, at 
small distances $r\rightarrow 0$, the coordinate $s$ stabilizes at 
the finite value $s_0$,
\be
s_0 = \left( {\rho \over 16} \right)^{1/5}
\ee
At this point the principal ${\bf S}^3$ orbit therefore collapses to an 
${\bf S}^2$ bolt at $r=0$, and the global topology of the surface $L$
can be identified with the spin 
bundle of ${\bf S}^2$ which we denote as $H^1$,
\be
L = H^1 \cong \mathbb{R}^2\times {\bf S}^2
\label{h1}\ee
For this smooth $L$, we may use the formulae \eqn{homo} to determine 
the homology of $X$, the M-theory lift. We see that $X$ is indeed 
described by a manifold of topology \eqn{myfriend} as advertised. 
In Section 4 we show explicitly that this $L$ coincides with the fixed 
point set of a suitable circle action on $X=\Sigma^-{\bf S}^4$. In 
this construction, the $Sp(1)$ action of Harvey and Lawson sweeps out the $Spin(7)$ manifold 
$X$ with a family of submanifolds. We will 
further show that the deformation parameter $\rho$, which measures the size 
of the ${\bf S}^2$ bolt of $L$, 
is related to the radius of the ${\bf S}^4$ bolt of $X$. Thus, the 
coassociative cone over the squashed three-sphere \eqn{lime} 
describes the reduction of the $Spin(7)$ cone over the squashed 
seven-sphere.

For this case, the D6-brane picture shows no sign of a 
geometrical transition to a 
manifold with different topology. Let us now turn to an example where 
such a transition does occur.

\subsection{D6-Branes and the Cone over $SU(3)/U(1)$}

We turn now to the main theme of the paper; the geometrical transition 
that occurs in the $Spin(7)$ cone over the 
weak $G_2$ holonomy Aloff-Wallach space, $N_{1,-1}=SU(3)/U(1)$. Let us start 
with the familiar resolution of this space to 
\be
\mathcal{Q} \cong \mathbb{R}^4 \times {\bf \mathbb{C}P}^2
\ee
The isometry group of this space is $U(3)$. A further symmetry arises from the 
$C$-field. The relevant cohomology groups are \cite{aw,GS}
\be 
H^2(\mathcal{Q};U(1)) = U(1)_J=H^2({N}_{1,-1};U(1))
\ee
which ensure that there is a single unbroken global symmetry, 
denoted $U(1)_J$, in the low-energy dynamics of M-theory 
compactified on $\mathcal{Q}$. 

\begin{figure}
\begin{center}
\epsfxsize=6in\leavevmode\epsfbox{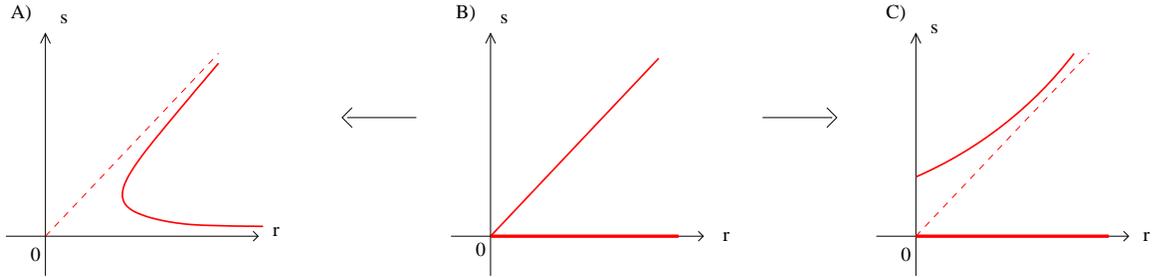}
\end{center}
\caption{A curve in the $r-s$ plane whose $Sp(1)$ orbits sweep 
out coassociative submanifolds: A) $L = {\bf S}^3 \times \mathbb{R}$, 
B) the singular cone, and C) $L=H^1\cup \mr^4$}
\label{hlfigb}
\end{figure}

Remarkably, the calibrated D6-brane locus which lifts to the manifold 
$\mq$ is described by the same Harvey-Lawson curve \eqn{curve} 
that we met in the previous section. To see this, consider the 
conical D6-brane described by the branch \eqn{lime}. To this 
we simply add a further flat D6-brane, lying on the locus
\be
s=0
\label{flatharry}\ee
The final configuration is depicted in Figure (\ref{hlfigb}B). 
These two coassociative submanifolds coincide at the conical singularity 
$s=r=0$. To resolve this singularity, we may deform the upper branch 
by $\rho\neq 0$ in the manner described in the previous section, while leaving 
the flat D6-brane described by \eqn{flatharry} unaffected. The resulting 
coassociative four-manifold $L$, shown in Figure (\ref{hlfigb}C), is 
simply the disjoint union
\be
L=H^1\cup\mr^4
\label{niceone}\ee
From equation \eqn{homo}, we see that this 
indeed has the requisite topology in order to lift to $\mathcal{Q}$. 
From the D6-brane perspective, the global $U(1)_J$ symmetry corresponds 
to the unbroken gauge symmetry on the flat D6-brane.
It is worth noting that this result is 
reminiscent of the D6-brane description of the $G_2$ 
holonomy manifolds with topology $\mr^3\times {\bf S}^4$ and 
$\mr^3\times\mcp^2$ \cite{aw,gt}. In this case, the addition of 
an extra flat D6-brane is also responsible for the difference between 
the ${\bf S}^4$ and $\mcp^2$ non-contractible cycle. 

\begin{figure}
\begin{center}
\epsfxsize=6in\leavevmode\epsfbox{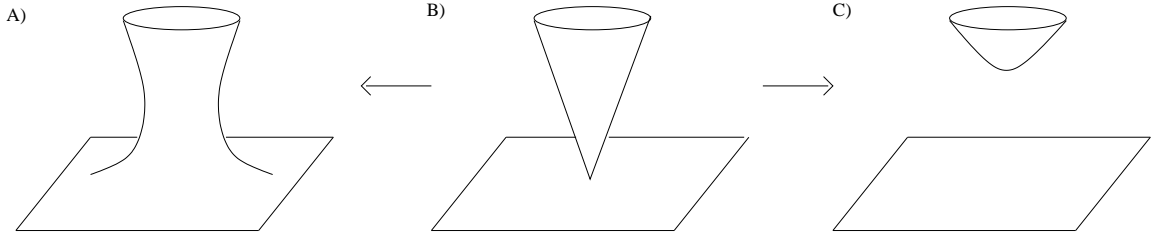}
\end{center}
\caption{The D6-brane in space-time, lying on the coassociative 
submanifolds: A) $L = {\bf S}^3 \times \mathbb{R}$, 
B) the singular cone, and C) $L=H^1\cup \mr^4$}
\label{itchyandscratchy}
\end{figure}

We are now in a position to see the geometrical transition from the 
D6-brane perspective. We simply 
note that there is a second resolution of the singular locus $L$ given 
by the union of \eqn{lime} and \eqn{flatharry}. 
This arises as another branch of the smooth curve \eqn{curve}, where
\be
0< s < {\sqrt{5} \over 2} r
\ee
In this situation, the two disjoint D6-branes smoothly join to lie 
on this branch as shown in Figure (\ref{hlfigb}A). The turning 
point of the curve occurs at $s=\ft12 r = (\rho/2^8)^{1/5}$.
Note that every point on the curve $s(r)$ is mapped into a 
three-sphere under the $Sp(1)$ action \eqn{HLaction}. Since this 
branch has both $s>0$ and $r>0$, there are no degenerate orbits and 
we conclude that the topology of the coassociative submanifold is 
given by
\be
L \cong {\bf S}^3 \times \mathbb{R}
\label{lfirstres}
\ee
This branch of $s(r)$ has two asymptotic components,
$s \sim {\sqrt{5} \over 2} r$ and $s \sim 0$, whose $Sp(1)$ orbits
coincide with the boundary of $H^1 \cup \mathbb{R}^4$. Taking the 
$\rho\rightarrow 0$ limit  returns us again to the singular 
description of two D6-branes. Figure (\ref{itchyandscratchy}) 
depicts a cartoon of the D6-brane transition in space-time.

From the homology relations \eqn{homo}, we see that a D6-brane 
placed on $L\cong {\bf S}^3 \times \mathbb{R}$ lifts to a 
$Spin(7)$ manifold of topology 
\be
X\cong \mathbb{R}^3\times {\bf S}^5
\ee
For this manifold $H^2(X;U(1))$ is trivial, ensuring that there are 
no symmetries associated with the C-field in
M-theory. This reflects the fact that there is no longer a flat 
D6-brane in the IIA picture. We conclude that the geometrical 
transition between coassociative submanifolds of topology 
$H^1\cup\mathbb{R}^4\ \leftrightarrow\ {\bf S}^3\times\mathbb{R}$ 
lifts in M-theory to a geometrical transition of $Spin(7)$ holonomy 
manifolds of topology 
$\mr^4\times\mcp^2\leftrightarrow \mr^3\times {\bf S}^5$.

We should point out that there are probably many other non-compact
$Spin(7)$ manifolds with conical singularities that arise
as M-theory lifts of D6-branes on coassociative submanifolds in $\mr^7$.
Thus, we already mentioned coassociative cones constructed
in \cite{Mashimo}. It would be interesting to study the geometry, 
and especially the physics, of these examples in more detail.

\subsection{Deformations of Coassociative Cones}

As we have argued, D6-branes on coassociative cones $L$ in $\mr^7$
lift to conical $Spin(7)$ manifolds in M-theory. In the following 
section, we will be interested in the dynamics of M-theory on such 
geometries. Therefore, it will be important to study their deformations 
and determine whether or not they are $L^2$ normalizable.
The latter aspect determines the interpretation in the low-energy
theory: $L^2$ normalizable deformations correspond to dynamical
fields, whereas non-normalizable deformations have infinite kinetic energy
and should rather be interpreted as true moduli, or coupling constants.

In M-theory on a non-compact (asymptotically conical) manifold $X$
deformations of the metric, $\delta g$, may be $L^2$-normalizable if 
the asymptotic conical metric is approached suitably quickly. 
A specific criterion can be obtained by
looking at the $L^2$ norm of $\delta g$ \cite{aw}:
\be
|\delta g|^2 = \int_X d^d x
\sqrt{g} g^{ii'} g^{jj'} \delta g_{ij} \delta g_{i'j'}
\ee
It follows that the deformation is $L^2$ normalizable if and only if
$\delta g / g$ goes to zero faster than $r^{-d/2}$, where $d$ is
the dimension of $X$. In other words, the critical exponent that
triggers $L^2$ normalizability is given by half the dimension of $X$.
In our first example, $X \cong \mr^4 \times {\bf S}^4$,
the explicit form of the metric is available, and one can directly 
check that the deformation corresponding to the change of the 
size of the ${\bf S}^4$ is not normalizable \cite{GS}
\be
|\delta g|^2 \to \infty
\label{nonnorm}
\ee
For the $Spin(7)$ manifolds $X\cong \Sfive$ and 
$\mq\cong \mr^4\times \mcp^2$, no explicit asymptotically 
conical metric is known. For this reason, we turn to the 
D6-brane locus to extract the relevant information. At finite 
string coupling, the D6-brane configuration lifts to an 
asymptotically locally conical $Spin(7)$ metric, which retains 
a finite asymptotic circle. The conditions for normalizability 
in such a metric are weaker than those of the corresponding 
asymptotically conical space. Hence, if the D6-brane deformation 
is non-normalizable, we can draw the same conclusion about the 
deformation of the $Spin(7)$ geometry. If the D6-brane locus is 
normalizable, no such conclusion may be reached.

The question of deformations of branes on non-compact, 
special Lagrangian submanifolds has been addressed 
recently by Neil Lambert \cite{Lambert}. For a submanifold $L$ of 
flat space, it was shown that the critical 
exponent that determines the $L^2$ normalizability
of the deformation is given by half the dimension of $L$. 
To see this, one must study the Lagrangian for the 
deformation modes \cite{Lambert}
\be
L_{eff} = \int_L d^p \sigma \sqrt{-{\rm det} (g + \delta g)}
- \int_L d^p \sigma \sqrt{-{\rm det} (g)}
\ee
where $g_{ij} = \eta_{ij} + \partial_i x^I \partial_j x^J \delta_{IJ}$
is the induced metric on the D6-brane, and $x^I$ are the D-brane 
embedding coordinates. 

Applied to the two examples of coassociative cones discussed above,
this criterion says that the deformation parameter $\rho$
in equation (\ref{curve}) is $L^2$-normalizable only if
\be
{\delta x \over \delta \rho} \sim r^{- \alpha}, \quad \alpha > 2
\ee
It is easy to show that this condition does not hold,
and therefore $\rho$ corresponds to a non-normalizable
deformation in our models.
In order to evaluate the derivative of $x^I$
with respect to the modulus $\rho$ it is convenient to gauge fix 
the asymptotic D6-brane world volume to be along $\mathbb{H}$
(this is precisely what we have for $X \cong \mr^3 \times {\bf S}^5$).
Since the distance to $\mathbb{H} \subset \mr^7$ is measured
by the radial variable $s$, we essentially need to evaluate
the asymptotic behavior of $ds / d \rho$.
{}From the defining polynomial (\ref{curve}) we compute:
\be
{ds \over d \rho} \times
\Big[ (4s^2 - 5r^2)^2 + 16 s^2 (4s^2 - 5r^2) \Big] = 1
\ee
and therefore,
\be
{ds \over d \rho} = \Big[ (\rho / s) + 16 s^{3/2} \rho^{1/2} \Big]^{-1}
\ee
which at large $r$ (large $s$) goes to zero like $r^{-3/2}$.
This proves that $\rho$ is indeed a non-normalizable deformation,
in agreement with (\ref{nonnorm}).

In a similar situation, it was conjectured
by Joyce \cite{Joyce} that deformations of (strongly asymptotically conical)
special Lagrangian cones $L \subset \mathbb{C}^3$ are, in fact, topological.
Namely, it was argued that the total number of deformations of $L$
is given by $b_1 (L) + b_0 (\partial L) -1$. This is to be compared
with deformations of {\it compact} special Lagrangian submanifolds,
parametrized by $b_1 (L)$ \cite{mclean}.
One might think that a similar topological formula holds for
deformations of coassociative submanifolds $L \subset \mr^7$,
in which case the number of deformations is likely to be given
by $b_2^{+} (L)$ (as in the compact case) plus some correction due
to the non-compactness of $L$.

\subsection{Anomalies and Supersymmetry}

The D6-brane configurations of the form $\mr^{1,2} \times L$, where $L$
is given by (\ref{h1}) or (\ref{niceone}), are, as they stand,
anomalous. In order to cancel the anomaly one must turn on a
half-integral flux of the gauge field strength $\mf$ through
the ${\bf S}^2$ bolt. 
Using the results of \cite{MMMS}, we now argue that this flux
does not break supersymmetry. 

As shown by Freed and Witten \cite{wittenfreed}, due to a global
anomaly for fundamental strings ending on type II D-branes, the
``$U(1)$ gauge field'' $\ma$ on a D-brane worldvolume $W$ should be
interpreted globally as a $\mathrm{spin}^c$ connection. This means
that, when $W$ is not a spin manifold, the quantization law for the 
field strength $\mf=d\ma$ is 
shifted from standard Dirac quantization. Specifically, for all
2-cycles $U_2\subset W$ we have
\be
\int_{U_2} {\mf \over 2 \pi} = {1 \over 2}\int_{U_2}w_2(W)\quad
\mathrm{mod}\ \mathbb{Z}\label{Fquant}
\ee
where $w_2(W)$ is the second Stiefel-Whitney class of $W$. 
Consider a D6-brane whose worldvolume has a component 
$\mr^{1,2}\times L$, where $L=H^1\cong \mr^2\times{\bf S}^2$. It is easy
to see that $H^1$ does not admit a spin structure, so that $w_2(L)\neq
0$. Consequently, in order to cancel the anomaly found by Freed and
Witten, one must turn on a half-integral flux of $\mf$ through ${\bf
S}^2\subset H^1$. Thus
\be
\int_{{\bf S}^2} \frac{\mf}{2\pi} \in \mathbb{Z}+\frac{1}{2}\label{fflux}
\ee
It is crucial that this flux does not break
supersymmetry. Fortunately, supersymmetric type II D-brane
configurations with non-zero gauge field strengths were studied in
\cite{MMMS}. The result relevant for us is that, for a D-brane wrapped 
over a coassociative submanifold, one can turn on a flux of the gauge
field without breaking supersymmetry, provided 
\be
\mf\cdot\Psi^{(3)}=0
\label{fpsi}\ee
For the explicit choice of $\Psi^{(3)}$ in equation \eqn{Psi}, 
we find that the flux is supersymmetric if and only if it is anti-self-dual. 
The two form dual to the ${\bf S}^2$ bolt of $H^1$ 
is indeed anti-self-dual (the self-intersection number of 
${\bf S}^2$ is -1), and we therefore do not break any supersymmetry by
turning on the flux (\ref{fflux}).

Let us now turn to the interpretation of this $\mf$-flux in the 
M-theory lift of these configurations. It was shown in \cite{GS} 
that the $\mf$-flux on the D6-brane may be identified with the 
$G$-flux in M-theory, using the relation
\be
H^4(X;\mathbb{Z})\cong H^2(L;\mathbb{Z})
\label{h2h4}\ee
Since the Freed-Witten anomaly requires the existence of $\mf$-flux, 
one may suspect that a similar consistency requirement leads to the 
presence of G-flux in the M-theory lift. Indeed, it was argued in 
\cite{GS} that the Freed-Witten anomaly for D6-branes wrapping a 
locus $L$ is equivalent to Witten's membrane anomaly for M-theory 
on the manifold $X$. Recall that the membrane 
path-integral is well-defined only if the G-field satisfies the 
shifted quantization 
condition \cite{witten1}
\be
a\equiv \left[\frac{G}{2\pi}\right]-\frac{\lambda}{2}\in 
H^4(X;\mathbb{Z})
\label{flux}\ee
where $\lambda(X)=p_1(X)/2 \in H^4(X;\mathbb{Z})$ is an integral class
for a spin manifold $X$. If $\lambda$ is even, one may consistently
set $G=0$. However, if  $\lambda$ is not divisible by two as an
element of $H^4(X;\mathbb{Z})$, one must turn on a half-integral 
$G$-flux in order to have a consistent vacuum. The relationship between 
type IIA string theory and M-theory then leads to an 
identification of $w_4(X)\cong \lambda(X)\ \mathrm{mod}\ 2$ and 
$w_2(L)$, under the mod 2 reduction of the isomorphism (\ref{h2h4}).

As explained in \cite{GS}, for both 
$X=\Sfour\cong \mr^4\times {\bf S}^4$ and $\mq\cong
\mr^4\times \mcp^2$ one can show that $\lambda(X)$ generates
$H^4(X;\mathbb{Z})\cong \mathbb{Z}$, and therefore one must turn on a
half-integral $G$-flux through the bolt ${\bf S}^4$ or $\mcp^2$,
respectively. 
Finiteness of the kinetic energy $\int G\wedge *G$, together with
the equations of motion, require $G$ to be an $L^2$-normalizable
harmonic 4-form on $X$. In the case of $X=\Sfour$, such a 4-form $G$
was constructed explicitly in \cite{pope}. We do not have the explicit asymptotically 
conical metric on $X=\mq$. However, the following
result of Segal and Selby \cite{SS} ensures the existence of an
$L^2$-normalizable harmonic 4-form $G$ on $X$ which represents the
generator of $H^4_{\mathrm{cpt}}(X;\mr)\cong \mr$. On a complete manifold $X$, an harmonic form
is necessarily closed and co-closed, so that $G$ defines a cohomology
class on $X$. In \cite{SS} it was argued that if the natural
map $f: H^p_{\mathrm{cpt}}(X)\mapsto H^p(X)$ takes a non-trivial
compactly supported cohomology class $b$ to a non-trivial ordinary cohomology
class $f(b)$, then there is a non-zero $L^2$-normalizable harmonic
$p$-form on $X$ representing $b$. For both $Spin(7)$ manifolds $X=\Sfour$
and $X=\mq$, the natural map 
$f:H^4_{\mathrm{cpt}}(X;\mathbb{Z})\rightarrow H^4(X;\mathbb{Z})$ 
maps the generator of
$H^4_{\mathrm{cpt}}(X;\mathbb{Z})\cong \mathbb{Z}$ to the generator of
$H^4(X;\mathbb{Z})\cong \mathbb{Z}$. In particular, the Thom class
which generates $H^4_{\mathrm{cpt}}(X)$ is represented by an
$L^2$-normalizable harmonic 4-form. Thus the existence of the 4-form
found explicitly in \cite{pope} is guaranteed by the general result of
\cite{SS}. We therefore set $G$ equal to the $L^2$-normalizable
harmonic 4-form predicted by \cite{SS}, appropriately normalized so
that $G$ satisfies the quantization condition (\ref{flux}).

We have seen previously that, from the D6-brane perspective, 
turning on a half-integral anti-self-dual $\mf$-flux through the $S^2$
bolt in $H^1$ does not break supersymmetry. Moreover, in M-theory,
this flux is dual to turning on a half-integral $G$-flux
through the ${\bf S}^4$ or $\mcp^2$ bolt of $\Sfour$ or $\mq$,
respectively. We choose conventions such that the parallel spinor 
of the $Spin(7)$ manifold has positive chirality (in the $\bf{8}_s$), ensuring that 
the Cayley form is self-dual\footnote{This agrees with the conventions 
of \cite{gary7} and \cite{adelag}, and allows comparison with the 
formulas for Calabi-Yau four-folds in \cite{GVW}. However, it is 
the opposite convention to \cite{pope}. Note also that the duality of 
the Cayley form is correlated with the use of D6-branes vs. 
anti-D6-branes in Sections 2.1 and 2.2.}. In these conventions, 
the $L^2$-normalizable 4-form constructed in \cite{pope} is self-dual.
We thus learn that the half-integral self-dual $G$-flux does not 
break supersymmetry. It would be interesting to derive this directly
from supersymmetry conditions in M-theory, extending a similar
analysis of fluxes on compact $Spin(7)$ manifolds \cite{adelag}. 

In this subsection we have seen that, in order to satisfy certain anomaly 
constraints, we must turn on background G-flux for M-theory on 
$\mq\cong\mr^4\times\mcp^2$. 
This is related to background $\mf$-flux for the D6-brane configurations 
of section 2.1 and 2.2. However, we have not yet determined which value 
of the G-flux arises after the conifold transition from $X=\Sfive$ to $\mq$.
For this, we must examine the boundary data more carefully.

\subsection{Flux at Infinity}

In order to define the problem of M-theory on an asymptotically conical 
$Spin(7)$ manifold, we must specify both the base of the cone $Y$, 
as well as details of the 3-form.  So far we have concentrated on 
the former. Here we turn our attention to the latter. 
We shall find that for vanishing asymptotic flux, M-theory on a $Spin(7)$  
manifold that is asymptotic to the cone on the Aloff-Wallach space 
$Y=N_{1,-1}$ has three branches\footnote{We thank E.~Witten for
explanations and very helpful discussions on these points.}.
Two of these branches correspond to the two choices of
sign for the half-integral $G$-flux through $\mcp^2 \subset \mq$, 
whereas the other branch corresponds to a change of topology to $\Sfive$.

As shown in \cite{GVW}, M-theory vacua which 
cannot be connected by domain walls are classified, in part, by 
$H^4(Y;\mathbb{Z})$. However, $H^4(Y;\mathbb{Z})$ 
is trivial for both $X=\Sfour$ and $\mq$ \cite{GS}. Therefore, 
the only asymptotic, non-geometric, data needed to specify the 
model is the value of the total flux at infinity \cite{GVW}
\be
\Phi_{\infty} =
N_{M2} + \frac{1}{192} \int_X \left(P_1^2 - 4 P_2\right)
+ {1 \over 2} \int_X {G \over 2 \pi} \wedge {G \over 2 \pi}
\label{anom}
\ee
Here $N_{M2}$ is the number of membranes filling
three-dimensional spacetime and, to preserve supersymmetry, we 
require $N_{M2}\geq 0$. The first and second Pontryagin forms of $X$ 
are, 
\be
P_1 = -\frac{1}{8\pi^2}\mathrm{tr}R^2\quad , \quad P_2 =
-\frac{1}{64\pi^4}\mathrm{tr}R^4 + \frac{1}{128\pi^4}(\mathrm{tr}R^2)^2
\ee
Note that anomaly
cancellation requires $\Phi_{\infty} =0$
for a compact space $X$ \cite{SVW,witten1}. Indeed, when $X$ is
compact, the $R^4$ terms give
\be
\frac{1}{192} \int_X \left(P_1^2 - 4 P_2\right) =
\frac{1}{192}\left(p_1(X)^2 - 4p_2(X)\right)
\ee
where $p_1^2(X)$ and $p_2(X)$ are Pontryagin numbers of $X$. When the
structure group of the tangent bundle of $X$ admits a reduction from
$Spin(8)$ to $Spin(7)$, one can show that $p_1^2-4p_2 = - 8\chi$,
where\footnote{The minus sign is correlated with our choice of 
orientation of $X$ by choosing the non-vanishing spinor field to be in
the $\bf{8}_s$ representation.} 
$\chi(X)$ is the Euler number of $X$. See, for example, \cite{IPW}. This
is equivalent to the existence of a nowhere vanishing spinor field on
$X$. We therefore get the usual anomaly term $\chi(X)/24$ familiar in
Calabi-Yau 4-fold compactifications.

When $X$ is non-compact, things are a little more complicated. It will
be crucial for us to compute the value of $\Phi_{\infty}$ for our
backgrounds, but we are 
presented with an immediate problem since, on a non-compact
manifold, the integral of the Pontryagin forms over $X$ is not a
topological invariant and we do not know explicitly the metric on
$X$. 

Fortunately, there is an effective way to compute the total flux at infinity,
provided that a dual D6-brane model is available. We have already provided 
one such dual picture in Section 2.2, consisting of D6-branes wrapping 
coassociative cycles in flat $\mr^7$. However, in this case the non-trivial 
part of the D6-brane worldvolume is also non-compact which does nothing to 
ameliorate our task. Thankfully, a second dual D6-brane model exists for the 
resolution of the cone to $\mq\cong\mr^4\times\mcp^2$. 
This was described in the introduction, and will be dealt with in 
greater detail in Section 4.  In this case, the D6-brane wraps the 
coassociative four-cycle $B=\mcp^2$ of the $G_2$ holonomy manifold 
$M^7 = \Lambda^2_+(\overline{\mcp}^2)$, the bundle of self-dual
\footnote{Note that we choose the orientation of the 
bolt to have $b^2_+=0$ and $b^2_-=1$. With this convention, deformations of the bolt are parameterized by $b^2_+=0$ 
\cite{mclean}. We thank B. Acharya for explanations of these
points. Throughout the rest of sections 2 and 3, we simply refer to the bolt as
$\mcp^2$, with the understanding that the opposite orientation
to the canonical one is to be used. However, one must be careful to note that the
bundle $\Lambda^2_+({\mcp}^2)$ is an \emph{entirely different}
bundle to $\Lambda^2_+(\overline{\mcp}^2)$, as one can see by comparing
Pontryagin classes. It is the total space of the latter bundle on
which the $G_2$ metric is defined.} 
two forms over $\overline{\mcp}^2$. This space has topology 
$M^7\cong\mr^3\times\mcp^2$.

The anomaly condition (\ref{anom}) relates the flux at infinity to the
number of space-filling membranes, the integral of the Pontryagin forms 
and the G-flux. After reduction to type IIA theory the effective membrane
charges become the effective charge of space-filling D2-branes.
What is the type IIA interpretation of the anomaly formula (\ref{anom})?

Since from the type IIA perspective the three-dimensional effective
theory is obtained by compactification on a seven-dimensional
$G_2$ manifold $M^7$, there is no contribution to the D2-brane
charge from the bulk.
However, in type IIA theory we also have a space-filling
D6-brane wrapped on the coassociative 4-cycle $B=\mcp^2$ inside $M^7$.
Due to the non-trivial embedding of the D6-brane worldvolume
in spacetime, the Ramond-Ramond fields in the bulk couple to the
gauge field strength $\mathcal{F}$ on the D6-brane. Specifically, we
have
\be
I_{\mathrm{WZ}}=-\int_{\mathbb{R}^3 \times B}
{C_*} \wedge \mathrm{ch}(\mathcal{F})
\wedge \sqrt{{\hat A (TB) \over \hat A (NB)}}
\label{anomwz}
\ee
where $TB$ (respectively $NB$) denotes the tangent
(respectively normal) bundle of $B=\mcp^2$ inside $M^7$,
and the Dirac genus $\hat A$ can be expressed in terms of
the Pontryagin forms as follows \cite{lawson}
\be
\hat A = 1 - {P_1 \over 24} + {7P_1^2 - 4 P_2 \over 5760} + \ldots
\ee
Comparing the $C_3$ coupling on the right-hand side
of (\ref{anomwz}) with the formula (\ref{anom}), we see that the type
IIA analogue of the latter is
\be
N_{D2} - \int_B  \sqrt{{\hat A (TB) / \hat A (NB)}} - 
\frac{1}{2}\int_B {\mf \over 2 \pi} \wedge {\mf \over 2 \pi}\label{2anom}
\ee
where $N_{D2}$ is the number of space-filling D2-branes. This is
naturally identified with $N_{M2}$ in M-theory. Recall that in the 
previous section we identified the (shifted) gauge field strength on the D6-brane 
with the (shifted) G-flux in M-theory, via the isomorphism (\ref{h2h4}). 
This suggests that the last terms in
(\ref{anom}) and (\ref{2anom}) are also naturally identified. 
Since a D6-brane wrapped on the bolt ${\mcp^2}\subset
M^7=\mr^3\times \mcp^2$ is anomalous unless one turns on a half-integral
flux of the gauge field strength \cite{wittenfreed}, we have 
\be
\int_{{\mcp^1}} \frac{\mf}{2\pi} = k \in \mathbb{Z}+\frac{1}{2}
\ee
which means the last term in (\ref{2anom}) takes the value 
$\ft12\int({\mf}/2\pi)^2=-k^2/2$. The
minus sign occurs since the orientation of $\overline{\mcp}^2$ in 
$M^7$ is such that it has $b^2_-=1$. The
contribution from the G-flux through $\mcp^2\subset\mq$ is
\be
\int_{{\mcp^2}} \frac{G}{2\pi} = k\label{kflux}\in \mathbb{Z}+\frac{1}{2}
\label{denmark}\ee
Using the fact that the self-intersection number of
$\mcp^2$ inside $\mq$ is equal to +1 in our conventions, we find that 
the last term of \eqn{anom} is given by $+k^2/2$. 
Finally, it is now natural to equate the remaining purely geometric terms,
so that
\be
\frac{1}{8} \int_X \left(P_1^2 - 4P_2\right) = \frac{1}{2} \int_{B} 
\Big( P_1(TB) - P_1(NB) \Big)
\label{chipp}
\ee
We should stress here that the right-hand side of this formula
is computed on a $G_2$ manifold $M^7$ and is manifestly a topological
invariant when $B$ is compact, whereas the left-hand side
is computed on the corresponding non-compact 8-manifold $X$ of $Spin(7)$
holonomy. Thus, we are able to compute the integral of the Pontryagin
forms by computing locally the two-brane
charge which is induced on the D6-branes. The formula (\ref{chipp})
may be proved in the compact case using a combination of $G$-index
theorems. Details will be presented elsewhere.

In \cite{GS} the Pontryagin classes in (\ref{chipp})
relevant for the $B$-picture for $X=\Sigma^-{\bf S}^4$ and
$X=\mq$ were computed. The right-hand side of (\ref{chipp}) is then given by
$((-4)-0)/2=-2$ and $((-3)-3)/2=-3$ respectively.
Remarkably, these are (minus) the topological Euler characters of $X$.

Putting these considerations together, we may now compute $\Phi_{\infty}$ 
for the case of $X=\mq$:
\be
\Phi_{\infty} = N_{M2} - \frac{3}{24} + \frac{1}{2}k^2
\label{phiq}\ee
where the half-integer $k$ determines the $G$-flux \eqn{denmark}. Thus, for 
$N_{M2}=0$ 
and the minimal value of $k=1/2$ or $k=-1/2$, we find $\Phi_{\infty} = 0$. 

Using this result for $\mq$,
together with some index theorems, we will now be able to
compute the integral of the $R^4$ terms for the $Spin(7)$ manifold
$\Sfive$. For zero $G$-flux, $G=0$, we shall again find that
$\Phi_{\infty}=0$ with $N_{M2}=0$.

The particular combination of Pontryagin forms of interest can be
written in terms of index densities for various elliptic operators on
$X$. Specifically, we have (see, for example, \cite{IPW})
\be
\frac{1}{16}\left(P_1^2 - 4P_2\right) = \ind \D_1 - \ind \D_- -
\frac{1}{2} \ind d\label{inddens}\ee

Here the index density for the exterior derivative $d$ is nothing but
the usual Euler density. The twisted Dirac operators $\D_1$ and $\D_-$
are the usual Dirac operators on $X$ coupled to the bundles
$\Lambda^1$ of 1-forms and $\Delta^-$ of negative chirality spinors
(in the $\bf{8}_c$).
The key observation is to note that the existence of a non-vanishing
spinor $\xi$ in the $\bf{8}_s$ representation provides us with a
natural isomorphism between the bundles $\Lambda^1$ and
$\Delta^-$. Explicitly, we can convert between the two using the formulae
\be
\psi = V \cdot \xi \quad  \quad  \quad V^{\mu} = \bar{\xi}\Gamma^{\mu}\psi\label{triality}
\ee
where $V$ and $\psi$ are an arbitrary (co)-vector and negative
chirality spinor, respectively.

For our supersymmetric compactification with $G$-flux, we have a
covariantly constant spinor $\xi$ in the $\bf{8}_s$, where the
derivative depends on the $G$-flux
\be
D_{\mu}(t)\xi \equiv \left(\nabla_{\mu}
-\frac{t}{288}G_{\rho\lambda\sigma\tau}\left(\Gamma_{\mu}^{\
\rho\lambda\sigma\tau}-8\delta_{\mu}^{\rho}\Gamma^{\lambda\sigma\tau}\right)\right)\xi=0
\ee
where we need to set $t=1$ in this equation. The usual covariant
derivative on $X$ is then given by setting $t=0$: $D(0)=\nabla$. We may now define twisted Dirac operators $\D_1(t)$ and
$\D_-(t)$. The index densities in (\ref{inddens}) are then for $\D_1(0)=\D_1$ and
$\D_-(0)=\D_-$, respectively.

The explicit form of the isomorphism (\ref{triality}) together with the fact that
$D(1)\xi=0$ implies that the spectra of the twisted Dirac operators
$\D_1(1)$ and $\D_-(1)$ are identical. In particular, the index of
these Dirac operators on $X$ with APS boundary conditions are
equal. Recall that the APS index theorem for a manifold with boundary
takes the following form
\be
\Index D = \int_X \ind D + \int_{\partial X} K - \frac{h+\eta(0)}{2}\label{APS}
\ee
Here $\ind D$ is the relevant index density for $D$, $K$ is a
boundary term depending on the second fundamental form, $h$ denotes the
multiplicity of the zero-eigenvalue of $D$ restricted to $\partial X$,
and $\eta$ is the usual APS function for the elliptic operator
$D$ restricted to $\partial X$. The boundary conditions are global; the projection onto the non-negative part of the spectrum on the
boundary is set to zero. 
We will write the APS theorem even
more schematically as
\be
\Index D = \int_X \ind D + \mathrm{boundary}\ \mathrm{terms}
\ee
where the boundary terms depend only on the boundary data. We will not
need to worry about the explicit form of these terms. 
It follows that we may write the integral of the Pontryagin forms as
\be
\frac{1}{16}\int_X \left(P_1^2 - 4P_2\right) = \Index \D_1(0) - \Index
\D_-(0) - \frac{1}{2} \chi(X) + \mathrm{boundary}\ \mathrm{terms}
\ee
where we have absorbed all of the boundary terms into the last
term. As we have already argued, 
\be
\Index \D_1(1) = \Index \D_-(1)\label{relation}
\ee
Now, importantly, the $G$-field vanishes at infinity. This was required
earlier for finiteness of the energy. Thus the restriction of
$\D_1(t)$ or $\D_-(t)$ to the boundary of $X$ is independent of $t$. In
other words, the spinor $\xi$ is a genuine Killing spinor on $Y$, even
in the presence of $G$-flux. We may now smoothly deform the operators
$\D_1(t)$, $\D_-(t)$ from $t=1$ back to $t=0$ without changing the
boundary data. Because of this last fact, the relation
(\ref{relation}) must continue to hold for all $t$: by (\ref{APS}) the index varies smoothly
under a smooth change of $D$ in the interior, and since the index is
an integer, it is therefore constant under such deformations. Notice that such arguments no
longer hold when the deformation is not smooth. For example, we can
deform the metric on $X=\mq$ whilst leaving the boundary data fixed by varying the size of the bolt
$\mcp^2$. This deformation is smooth, except when we pass the conifold
point. In fact, the index jumps as we move from $\mq$ to $\Sfive$, as
we shall see presently. 


It follows that for a $Spin(7)$ manifold $X$ we have 
\be
\frac{1}{16}\int_X \left(P_1^2 - 4P_2\right) = -\frac{1}{2} \chi(X) +
\mathrm{boundary}\ \mathrm{terms}
\label{pont}\ee
We have just argued, using duality, that 
for $X=\mq$ the sum of all the boundary terms must vanish, since the left 
hand side of the relation (\ref{pont}) gave precisely the topological 
result $-\chi(X)/2$. For $X=\Sfive$, the Euler class vanishes since the 
manifold is contractible to ${\bf S}^5$. 
Moreover, the boundary terms are equal to the boundary terms
for $\mq$ since both manifolds have the same asymptotics. But we have
just argued that the sum of the boundary terms is zero. Thus, for
$X=\Sfive$ with $N_{M2}=0$ and $G=0$, we have once again $\Phi_{\infty}=0$.

There is a simple physical explanation of this result. For $G=0$ and
no space-filling M2-branes, the only contribution to the total flux
(\ref{anom}) is from the $R^4$ terms. There exists a type IIA dual for
M-theory on $X=\Sfive$ which involves RR 2-form flux on a $G_2$
manifold with no D6-branes. This was described briefly in the introduction 
and will be dealt with in detail in Section 4. In the absence of 
space-filling D2-branes,
there is no other contribution to the effective D2-brane charge in
this type IIA string theory configuration. Thus the $R^4$ terms in
M-theory must vanish.


It is satisfying that this combination of physical and
mathematical arguments is self-consistent.

\subsection{Parity Transformations}

Let us quickly recap: 
we have examined M-theory on the background which asymptotes to the 
$Spin(7)$ cone over 
the Aloff-Wallach space $N_{1,-1}$, with vanishing flux at infinity. 
We have shown that there are three choices of 
supersymmetric vacua satisfying this boundary data, and therefore 
three possible branches for the moduli space of M-theory on
this conical singularity. One branch consists of $X=\Sfive$ with $G=0$. 
The other two branches correspond to $X=\mq\cong\mr^4\times\mcp^2$ 
with either plus or minus 
a half unit of $G$-flux through the $\mcp^2$ bolt. 

It is natural to wonder how, if at all, these three branches join 
together. In Section 2.2, we presented evidence that one may move 
from a branch with topology $X=\Sfive$ to a branch of topology 
$\mq$. But which sign of G-flux does this 
transition choose? Our conjecture here is that all three branches 
meet at a singular point of moduli space. To see that this must 
be the case, it is instructive to study the action of parity on 
each of these backgrounds. 

In odd space-time dimensions, parity acts by inverting an odd number 
of spatial coordinates. In the present case of M-theory compactified 
on $\mr^{1,2}\times X$, it is convenient to take the action of the 
parity operator to be
\be
P: x^1\ \rightarrow\ -x^1
\label{parity}\ee
for $x^1\in\mr^{1,2}$, with all other coordinates left invariant. The 
advantage of such a choice is that parity in M-theory coincides 
with parity in the low-energy three-dimensional theory. With the 
natural decomposition of gamma matrices into a $3-8$ split, it is 
simple to show that this remains true for the fermions, with an 
eleven dimensional Majorana fermion $\epsilon$ transforming as
\be
P: \epsilon \rightarrow \Gamma_1\epsilon
\ee
In order for the Chern-Simons interaction $C\wedge G\wedge G$ of eleven 
dimensional supergravity 
to preserve parity, we must also choose the action on the 3-form field 
of M-theory,
\be
P:\left\{\begin{array}{ll} C_{ijk}(x)\ \rightarrow\ +\,C_{ijk}(Px) & 
\mbox{\ \ if\ $i,j$\ or\ $k=1$} \\ 
C_{ijk}(x)\ \rightarrow\ -\,C_{ijk}(Px) & 
\mbox{\ \ if\ $i,j$\ and\ $k\neq 1$} \end{array}\right.
\label{eng3den0}\ee 
While this ensures that M-theory respects parity, certain 
backgrounds may spontaneously break this symmetry. This is not the 
case for the geometry $X\cong \Sfive$. However, the requirement of 
non-vanishing G-flux in the geometry $\mq$ ensures that parity is 
indeed broken in this background, since
\be
P:\int_{\mcp^2}\frac{G}{2\pi}\ \rightarrow\ -\int_{\mcp^2}\frac{G}{2\pi}
\ee
Thus we see that the two branches with $G=\pm\ft12$ transform into each 
other under a parity transformation. It follows that, if either 
of these branches can be reached from the parity even branch with 
$X=\Sfive$, then both may be reached. In the following section we 
shall argue that this may be understood via the condensation of a 
parity odd state. 
 
Notice that similar comments apply to the $Spin(7)$ manifold 
$\mathbb{B}_8\cong\mr^4\times{\bf S}^4$. Once again, the 
presence of G-flux ensures that parity is spontaneously broken. 
In this case however, there is no third parity invariant branch, 
allowing for the possibility that the two branches with flux 
$\pm\ft12$ are disjoint. However, given the similarity between 
the D6-brane pictures, described in Sections 
2.1 and 2.2, it seems likely that the branches are once again connected.


\section{Low-Energy Dynamics and Condensation of Five-Branes}

In this section we would like to consider the low-energy
dynamics of M-theory compactified on a $Spin(7)$
manifold to three-dimensions with ${\cal N}=1$ (two supercharges)
supersymmetry. To illustrate our methods
we will first consider the similar, but simpler, example
of IIA string theory compactified to three
dimensions on a manifold of $G_2$ holonomy. The dynamics of the 
massless modes were discussed in detail by Atiyah and Witten 
\cite{aw}. Here we include the effects from massive wrapped branes 
which become light at certain points of the moduli space, and 
rederive some of the results of Atiyah and Witten \cite{aw}
in a new fashion. We will then apply the lessons learnt to the
$Spin(7)$ case.

\subsection{Type II Strings on Manifolds of $G_2$ Holonomy}

\subsubsection{The Cone over $\mcp^3$}

The cone over $Y={\mathbb C}{\bf P}^3$ is the first, and simplest,
example considered by Atiyah and Witten \cite{aw}.
The singularity may be resolved to a manifold of topology 
\be
X\cong{\mathbb R}^3\times {\bf S}^4
\ee
Let us consider IIA string theory compactified on $X$ to 
$d=2+1$ dimensions. This preserves ${\cal N}=2$ supersymmetry. 
Following \cite{aw}, we firstly examine the massless Kaluza-Klein 
modes. The volume of the ${\bf S}^4$ cycle provides a real scalar 
field,
\be
\phi={\rm Vol}({\bf S}^4)
\ee
Importantly, this deformation is normalizable and $\phi$ is
dynamical \cite{aw}. The ${\cal N}=2$ supersymmetry requires that
$\phi$ be accompanied by a further scalar field arising from the
harmonic $L^2$-normalizable three-form $\omega$ such that
\be
\int_{{\mathbb R}^3}\omega\neq 0
\ee
where ${\mathbb R}^3$ is the fiber over ${\bf S}^4$. The existence of
$\omega$ ensures that the RR three-form $C_3$ has a zero mode
\be
C_3=\sigma \omega
\ee
and $\sigma$ provides the second massless scalar in $\mr^{1,2}$. 
Large gauge transformations of $C_3$ mean that $\sigma$ is a
periodic scalar. Moreover, the parity transformation \eqn{eng3den0} 
ensures that it is actually a pseudoscalar. 
Atiyah and Witten \cite{aw} combine $\phi$ and
$\sigma$ into the complex field $\phi\exp(i\sigma)$,
which is the lowest component of a chiral multiplet. Here
we choose instead to dualise the scalar $\sigma$ for a 
three-dimensional $U(1)$ gauge field,
\be
d\sigma={}^\star dA
\ee
where the Hodge dual $\star$-operator is defined on the 
${\mathbb R}^{1,2}$ Minkowski space 
transverse to $X$. In this language, the
low-energy dynamics is defined in terms of a $U(1)$ vector multiplet.
To see the utility
of this duality, let us consider the effect of D4-branes wrapping
the coassociative ${\bf S}^4$. The lowest mass state occurs if the 
field strength on the D4-brane is set to zero. For $\phi\gg 0$, 
the real mass of the D4-brane is given by
\be
M_{D4}\sim\phi+\frac{1}{48}\int_{{\bf S}^4} \Big( p_1(T{\bf S}^4)
-p_1(N{\bf S}^4) \Big) \sim \phi-\phi_0
\nn\ee
where the fractional D0-brane charge  
induced by the the curvature couplings \eqn{anomwz} was calculated 
in Section 2.5 and contributes a 
negative, bare, real mass $\phi_0=1/12$.  
(See \cite{clifford} for a nice discussion). Since the mass is 
proportional to $\phi$, 
supersymmetry requires that this state is charged under the 
$U(1)$ gauge field. 
To see this explicitly, note that the charge of these
states is measured by the asymptotic RR-flux,
\be
\int_{{\mathbb R}^3\times {\bf S}^1} dC_3 =
\int_{{\bf S}^1}d\sigma
\label{eng17bra0}\ee
where ${\mathbb R}^3$ is the fiber of $X$ and ${\bf S}^1$
is a large space-like circle in ${\mathbb R}^{1,2}$ surrounding
the point-like D4-brane. The D4-brane is therefore a global vortex
in $\sigma$ or, alternatively, is charged electrically
under $A$. Notice that the state corresponding to a single
D4-brane is not in the spectrum since it has logarithmically
divergent mass. (The same is true for the M5-brane wrapping
${\bf S}^4$ in M-theory which leads to a BPS string in the four
dimensional effective theory). Nevertheless, its effects are
still important.

How does this D4-brane appear in the low-energy effective action? 
The simplest hypothesis is that a correct quantization of the 
D4-brane wrapped around the calibrated ${\bf S}^4$ yields a 
single, short (BPS) chiral multiplet. To see that this is 
consistent, let us examine the global symmetries of the model.
Recall that these arise from both geometrical isometries of $X$, as
well as gauge symmetries of the $C_3$ field \cite{aw}. The
latter are more important for us. The symmetries are determined
by large gauge transformations at infinity, while those which
can be continued into the interior of $X$ are unbroken symmetries.
We have \cite{aw},
\be
H^2(Y;U(1))=U(1)_J\quad ,\quad H^2(X;U(1))=0
\ee
The model therefore has a single, broken, $U(1)$ global symmetry
which acts on the dual photon as $\sigma\rightarrow\sigma +c$.
Note in particular that there are no further flavor symmetries.
While this does not rule out the existence of further matter multiplets,
constrained by a suitable superpotential, it does suggest that the simplest
possibility is to have a single chiral multiplet, which we denote 
as $q$. 

However, this is not the full story since, in three-dimensions, 
massive charged particles do not necessarily decouple from the 
low-energy dynamics. Rather, they lead to the generation of 
Chern-Simons couplings. Since we have ``integrated in'' the 
D4-brane state $q$ to describe our effective theory, we must 
compensate by the introduction of a bare Chern-Simons coupling 
in our theory. This is such that, upon integrating out the 
massive D4-brane, the effective Chern-Simons coupling vanishes. 
To determine this Chern-Simons coupling, we need both the 
charge of the chiral multiplet $q$, and the sign of the mass of 
the fermions. The former is determined by \eqn{eng17bra0} to 
be $+1$, while the latter may be fixed, by convention, to be 
positive.
Thus, integrating in the fermions associated to $q$ 
gives rise to a bare Chern-Simons parameter  $\kappa= -\ft12$. 
We are thus led to the simplest hypothesis for the matter content; 
a Maxwell-Chern-Simons theory, with single charged 
chiral multiplet. 

\begin{figure}
\setlength{\unitlength}{1.25em}
\begin{center}
\begin{picture}(23,5)
\qbezier(15,4)(2,2)(15,0)
\qbezier(15,4)(14,2)(15,0)
\qbezier(15,4)(16,2)(15,0)
\put(17,2){\vector(1,0){2}}
\put(19.2,2){$\phi$}
\end{picture}\end{center}
\caption{The quantum moduli space of $\mathcal{N}=2$
three-dimensional Chern-Simons-Maxwell theory with a 
single chiral multiplet.}
\label{figyes}
\end{figure}
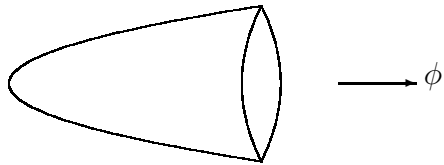

The supersymmetric completion of the Chern-Simons term includes 
a D-term coupling to $\phi$. Physically, this can be understood 
as arising from integrating in the complex scalar field $q$. 
Thus, the potential energy of the low-energy dynamics 
is given by,
\be
V=e^2(|q|^2-\kappa\phi)^2+(\phi-\phi_0)^2|q|^2
\label{dterm1}\ee
where $e^2$ is the gauge coupling constant. Naively, this theory has 
no moduli space of vacua. However, if we set $\phi >\phi_0$ 
then, upon integrating out the chiral multiplet, the renormalized 
$\kappa$ vanishes and we see that this is indeed a supersymmetric 
vacuum state. The moduli space of this theory is thus given by 
$\phi>\phi_0$, over which the dual photon is fibered, as shown in 
Figure (\ref{figyes}). 
This fiber degenerates at $\phi=1/12$ where the chiral multiplet is massless, 
and $U(1)_J$ is restored at this point. Atiyah and Witten argue that this 
point is smooth \cite{aw}.

Let us comment briefly on parity. Type IIA string theory on 
the $G_2$ manifold $X\cong\mr^3\times {\bf S}^4$ is parity invariant. 
The same is true of our low-energy description, arising after 
integrating out the D4-brane. Our effective theory also 
includes higher dimensional parity breaking operators, for example 
$(A\wedge F)^3$. However, as usual, our theory is simply not to 
be trusted at such scales since we have ignored many other 
contributions. 

It is curious to note that the smooth moduli space depicted in (\ref{figyes})
may be embedded as the ${\bf S}^1$ fiber of Taub-NUT space.
To see this, consider ${\cal N}=4$ supersymmetric SQED with a 
single hypermultiplet, in three dimensions. It is well known that the 
four-dimensional Coulomb branch of this model is endowed with 
the smooth hyperk\"ahler Taub-NUT metric, arising at one-loop \cite{seiberg} 
\be
ds^2 = \left(\frac{1}{e^2}+\frac{1}{\phi}\right)(d\phi^2 +\phi^2
d\theta^2 + \phi^2\sin^2\theta d\lambda^2)+\frac{1}{4}
\left(\frac{1}{e^2}+\frac{1}{\phi}\right)^{-1}(d\sigma+ \cos\theta d\lambda)^2
\ee 
The ${\cal N}=4$ gauge theory has a $SU(2)_N\times SU(2)_R$ R-symmetry 
group. 
Of these, only the former acts as an isometry on the Coulomb branch, 
rotating the three complex structures. The Coulomb branch has a further 
tri-holomorphic $U(1)_J$ isometry, which rotates the dual photon $\sigma$. 

One may flow from this ${\cal N}=4$ gauge theory to the ${\cal N}=2$ 
gauge theory of interest by turning on relevant, supersymmetry breaking, 
operators. These operators may be conveniently introduced by weakly 
gauging the diagonal global symmetry 
$U(1)_D\subset U(1)_N\times U(1)_R \times U(1)_J\subset SU(2)_N\times 
SU(2)_R\times U(1)_J$ \cite{tong}. This gives masses to precisely 
half of the fields as required. Moreover, integrating out half of the 
hypermultiplet gives rise to the bare Chern-Simons coupling 
$\kappa=-\ft12$. However, because the relevant deformation is 
associated to a symmetry, we may follow it to the infra-red, where it 
may be understood as the generation of a potential on the moduli space 
of vacua, proportional to the length of the Killing vector corresponding 
to simultaneous rotations of $\sigma$ and $\lambda$, 
\be
V=\phi^2\sin^2\theta\left(\frac{1}{e^2}+\frac{1}{\phi}\right)+
\frac{1}{4}\left(\frac{1}{e^2}+\frac{1}{\phi}\right)^{-1}(1+\cos\theta)^2
\ee
This potential vanishes on the two-dimensional submanifold $\theta=0$. 
This submanifold is precisely the moduli space of Figure (\ref{figyes}).

\subsubsection{The Cone over $SU(3)/U(1)^2$}

The next example we consider is the cone over 
$Y=SU(3)/U(1)^2$
which was also discussed by Atiyah and Witten \cite{aw}.
The conical singularity may be resolved to a manifold of topology
\be
X\cong{\mathbb R}^3\times{\mathbb C}{\bf P}^2
\ee 
The story is
similar to that above. Once again,
the volume of the four-cycle yields a dynamical real scalar,
$\phi={\rm Vol}({\mathbb C}{\bf P}^2)>0$, and a normalizable
harmonic 3-form provides the supersymmetric partner
$\sigma$ \cite{aw}. This latter, periodic, pseudoscalar is 
dualized in favor of
a $U(1)$ gauge field. The only question is what
charged matter arises from a D4-brane wrapped on
${\mathbb C}{\bf P}^2$. This time the analysis is somewhat
different. Although ${\mathbb C}{\bf P}^2$ is a supersymmetric
cycle, the lack of spin structure implies that we cannot wrap
a D4-brane on it without including suitable world-volume field
strengths. These require fluxes,
\be
\int_{{\mathbb C}{\bf P}^1} \frac{\mf}{2\pi}  = k \in {\mathbb Z}+\ft12
\ee
For supersymmetry \cite{MMMS}, the flux must satisfy \eqn{fpsi}.
In our conventions, this requires that $\mf$ is anti-self-dual which, 
since the bolt $B=\overline{\mcp}^2$ has $b_2^+ (B) =0$ and 
$b_2^-(B)=1$, is indeed the case. 
In our analysis, we wish to include only the states which become 
light in some regime of moduli space. From the semiclassical 
mass formula, for $\phi\gg 0$ we have 
\be
M_{D4} \sim \phi - 
{1 \over 2} \int_{\mcp^2} \frac{\mf}{2\pi} \wedge \frac{\mf}{2\pi}
+ {1 \over 48}\int_{\mcp^2} \Big( p_1 (T{\mcp^2}) - p_1(N{\mcp^2} ) \Big) 
\ge \phi
\label{atlast}\ee
The last term was calculated in Section 2.5 (see the paragraph above 
\eqn{phiq}), and yields a negative contribution to the mass; 
$-\phi_0=-1/8$. Nevertheless, 
the fact that the anomaly cancellation requires a non-zero, 
anti-self-dual, flux ensures that the inequality \eqn{atlast} holds, 
and is saturated only by the two states with minimal flux, 
$k=\pm 1/2$. All states with larger values of flux remain 
massive throughout moduli space. We therefore include in the 
low-energy description only the two states with minimal flux.

What quantum numbers do these two states carry in the low-energy 
effective theory? 
From \eqn{eng17bra0}, a D4-brane with either sign of flux has 
charge $+1$ under the $U(1)$ gauge group. However, the low-energy 
fermion fields have equal, but opposite, mass. To see this, note 
that under parity we have $k\rightarrow -k$. This follows from the 
fact that $\mf$ transforms in the same way as the NS-NS 2-form field, 
whose own transformation properties may be deduced from \eqn{eng3den0}. 
Since the action of parity in IIA string theory coincides with the 
action in the low-energy three-dimensional theory\footnote{It also 
acts as worldsheet parity for the IIA string.}, the signs of the 
fermion masses are reversed. It is conventional in ${\cal N}=2$, 
$d=2+1$ theories to adjust the complex structure of chiral multiplets 
so that the sign of the gauge field charge coincides with the 
sign of the fermion mass. 
We therefore find the low-energy effective dynamics to be governed 
by a three dimensional, ${\cal N}=2$ $U(1)$ gauge theory with two 
chiral multiplets $q$ and $\tilde{q}$ of charges $+1$ and 
$-1$ respectively and equal, but opposite, fermion masses. 
Unlike the previous case, the ``integrating in'' 
of two chiral multiplets with opposite fermion masses means that no bare 
Chern-Simons term is generated.

The presence of two, minimally coupled, chiral multiplets endows the 
low-energy dynamics with a flavor symmetry, 
under which both $q$ and $\tilde{q}$ transform with the same charge. 
This symmetry may be seen from M-theory, where the C-field 
yields the following global symmetries \cite{aw}:
\be
H^2(Y;U(1))=U(1)_J\times U(1)_F \quad,\quad H^2(X;U(1))=U(1)_F
\ee
The interpretation of this is that there exist two $U(1)$ 
global symmetries,
one of which, $U(1)_J$, is spontaneously broken. We have
suggestively labelled the unbroken symmetry $U(1)_F$. To
see that it is indeed the above flavor symmetry, it suffices
to note that the corresponding, non-dynamical, gauge potential
is given by
\be
A_F=\int_{{\mathbb C}{\bf P}^1} C_3
\ee
where $\mcp^1\subset\mcp^2$. This ensures that a D4-brane with 
flux indeed carries the requisite charge. 

IIA string theory on this background also contains an object 
arising from the D2-brane wrapping the $\mcp^1\subset\mcp^2$. From 
the above discussion, we learn that this state carries flavor, 
but no gauge, charge. In fact, it is simple to construct this 
state in the low-energy theory. Its identity follows 
from the observation that it may be constructed from a D4-brane 
with $k=+1/2$ bound to an anti-D4-brane also with $k=+1/2$. 
This relates the D2-brane to the $q\tilde{q}$ bound state, 
which indeed carries the correct quantum numbers.  At first sight, 
there appears to be a contradiction. 
The D2-brane is not wrapped on a calibrated cycle, and does 
not therefore give rise to a BPS state. In contrast, the 
operator $q\tilde{q}$ is holomorphic. However, it is a dynamical 
question whether the bound state of $q$ and $\tilde{q}$ saturates 
the BPS bound and, using mirror symmetry, one may argue that it 
does not. To see this, recall that this state is dual to a vortex 
state on the Higgs branch of a theory with two oppositely charged 
chiral multiplets \cite{ahiss}. But no classical BPS vortex solution 
exists in this theory\footnote{This follows from the fact that a line 
bundle of negative degree cannot have a non-zero holomorphic section.}.

\begin{figure}

\setlength{\unitlength}{0.9em}
\begin{center}
\begin{picture}(22,11)

\put(10,4){\line(1,2){3}}
\put(10,4){\line(-1,2){3}}
\qbezier(7,10)(10,12)(13,10)
\qbezier(7,10)(10,8)(13,10)
\put(14,9){Higgs}

\qbezier(0,6)(5,6)(10,4)
\qbezier(0,2)(5,2)(10,4)
\qbezier(0,6)(1,4)(0,2)
\qbezier(0,6)(-1,4)(0,2)
\put(3,4){$v_-$}\put(0,0.3){Coulomb}

\qbezier(20,6)(15,6)(10,4)
\qbezier(20,2)(15,2)(10,4)
\qbezier(20,6)(21,4)(20,2)
\qbezier(20,6)(19,4)(20,2)
\put(16,4){$v_+$}\put(18,0.3){Coulomb}

\end{picture}\end{center}
\caption{The quantum moduli space of $\mathcal{N}=2$
three-dimensional SQED with two chiral multiplets.}
\label{figk}
\end{figure}
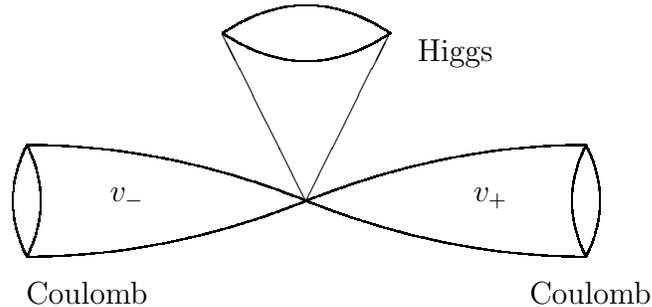

Finally, let us turn to the moduli space of this theory, and ask 
what happens as $\phi\rightarrow 0$. The classical 
scalar potential is given by,
\be
V=e^2(|q|^2-|\tilde{q}|^2)^2+\phi^2(|q|^2+|\tilde{q}|^2)
\ee
So far we have restricted attention to the Coulomb branch, 
with $\phi\neq 0$. However, for $\phi=0$, there exists 
a one complex dimensional Higgs branch in which $U(1)_F$ 
is spontaneously broken, and $U(1)_J$ is unbroken. The 
quantum dynamics of this three dimensional gauge theory were 
examined in \cite{ahiss,berk}. Here it is shown that the 
Coulomb branch bifurcates into two cigar-shaped branches, joined 
together at $\phi=0$ where they meet the Higgs branch. 
It is thought that at the junction of the three branches
there lives an interacting superconformal field theory. 
The two Coulomb branches
are parameterized asymptotically by $v_\pm=\exp(\pm\phi\pm i\sigma)$. 
The final quantum moduli space
is sketched in Figure \ref{figk}. The authors of \cite{ahiss}
further conjecture that, at strong coupling, the theory enjoys a
triality symmetry which interchanges the two Coulomb branches and
the Higgs branch. They argue that the physics is thus dual to the
Landau-Ginzburg model with three chiral multiplets,
$\Phi_i$, $i=1,2,3$ and the superpotential
\be
{\cal W}=\Phi_1\Phi_2\Phi_3
\ee
This is in perfect agreement with the results of \cite{aw}, where
the existence of three branches was deduced using a discrete
$S_3$ symmetry group of the $G_2$ holonomy manifold $X$. Each of 
the three branches corresponds to a manifold of topology 
$X\cong\mr^3\times\mcp^2$, with only the unbroken $U(1)$ 
symmetry group distinguishing them. 

In \cite{gt} it was shown that $X$ has a non-normalizable 
deformation which preserves both homology and holonomy. There are three
such ways to perform this deformation, each of which
preserves only one of the three branches of moduli space. Let
us see how to capture this behavior from the perspective
of the low-energy dynamics. Since both supersymmetries
and global symmetries of the low-energy dynamics are left intact,
this deformation can correspond to only two possible parameters;
a real mass parameter $m$ or a FI parameter $\zeta$. Including
both, the scalar potential reads,
\be
V=e^2(|q|^2-|\tilde{q}|^2-\zeta)^2+(\phi+m)^2|q|^2+(-\phi+m)^2
|\tilde{q}|^2
\label{dterm2}\ee
There are indeed precisely three such combinations of these parameters
which preserve a given branch of the vacuum moduli space
\ba
 \zeta\neq 0\ ,\ m=0 && {\rm Higgs\ survives} \nn\\
\zeta=-m\neq 0 && v_+\ {\rm Coulomb\ survives} \nn\\
\zeta=+m\neq 0 && v_-\ {\rm Coulomb\ survives} \nn
\ea

\subsection{M-theory on the $Spin(7)$ Cone over $SU(3)/U(1)$}

We now turn to the main topic of this paper: the dynamics of M-theory 
compactified on the cone over $Y=SU(3)/U(1)$. We will use the intuition 
gleaned from the previous sections to provide a consistent 
picture of the topology changing transition from $X\cong\Sfive$ to 
$\mq\cong{\mathbb R}^4\times {\mathbb C}{\bf P}^2$.

Let us start with M-theory compactified on 
$X\cong{\mathbb R}^3\times{\bf S}^5$. As for the $G_2$ examples 
discussed in the previous subsection, the volume of the ${\bf S}^5$ 
yields a real parameter,
\be
\phi={\rm Vol} ({\bf S}^5)
\ee
However, there is an important difference with the $G_2$ holonomy 
examples discussed in the previous section. As shown in Section 2.3, 
using the relationship to coassociative cones, fluctuations of 
$\phi$ are non-normalizable. Therefore $\phi$ plays the role of 
a modulus in the low-energy dynamics. Of course, one could imagine 
compactifying the $Spin(7)$ manifold, with the geometry $\Sfive$ 
providing a good description in the neighborhood of a conical 
singularity, in which case $\phi$ is once again promoted to a 
dynamical field. 

We now turn to the massless modes arising from the 
C-field. Although no explicit harmonic 3-form (or, indeed, 
a metric!) is known on $\Sfive$, there is a simple argument to 
ensure the existence of such an object. To see this, 
consider the symmetries of M-theory
on $\Sfive$. The global symmetry at infinity arising from the
$C$-field is given by
\be
H^2(Y;U(1)) \cong U(1)_J
\ee
where the generator is dual to the ${\bf S}^2$ fiber of $Y\rightarrow
{\bf S}^5$. This $U(1)$ symmetry is spontaneously broken in the
interior
\be
H^2(X;U(1)) \cong 0
\ee
Thus, in the conical limit $\phi=0$, where the ${\bf S}^5$ bolt
collapses to zero size, the low-energy theory has a global $U(1)_J$
symmetry. This is spontaneously broken for $\phi>0$, with
\be
\delta C = d\Lambda
\ee
where $\Lambda$ is a 2-form dual to the ${\bf S}^2$ fiber of
$Y\rightarrow {\bf S}^5$. Since $U(1)_J$ acts non-trivially 
on the low-energy theory, it must give rise to a Goldstone mode,
\be
C = \sigma \omega_3
\ee
predicting the existence of an harmonic 3-form $\omega_3$ 
which represents the generator of the
compactly supported cohomology $H^3_{\mathrm{cpt}}(X)\cong
\mr$.

There remains the question of whether this 3-form is
$L^2$-normalizable, or, equivalently, whether the periodic 
pseudoscalar $\sigma$ is dynamical. This remains an open problem. 
We will denote normalization of the kinetic term for $\sigma$ as 
$e^2$; the non-normalizable limit corresponds to $e^2\rightarrow \infty$. 
As in previous examples, we dualise $\sigma$ in favor of a $U(1)$ 
gauge potential $A$. In terms of these new variables, the field 
strength kinetic term is normalized as the usual $1/e^2$. Note that 
if $\omega_3$ is non-normalizable, we are dealing with the strong coupling 
limit of the gauge theory. 

As in the $G_2$-holonomy examples of the previous section, 
extra massive states arise from wrapped branes. In the present 
case, these come from M5-branes wrapping $\Sfive$. In the 
semi-classical limit $\phi \gg 0$, these give rise to states 
of real mass $\phi$, charged under the $U(1)$ gauge field. 
From the perspective of the coassociative 
D6-brane locus $L$ of Figure (\ref{hlfigb}A), this M5-brane corresponds 
to a D4-brane with topology of a four-disc $D^4$, whose boundary 
${\bf S}^3$ wraps the minimal volume three-cycle of $L$. 
What type of matter does quantization of
the wrapped M5-brane yield? As the ${\bf S}^5$ is not a
calibrated submanifold, we again do not expect a supersymmetric
multiplet. But, since $\mathcal{N}=1$ supersymmetry in
three dimensions does not admit BPS particle states, this is no 
great limitation.  The simplest possible matter content, consistent 
with the symmetries of the theory, occurs if the M5-brane gives 
rise to a single complex scalar  multiplet $q$. We assume that 
this is the case and that, as in \cite{Strominger,gms}, this is 
the only single particle state to become light as 
$\phi\rightarrow 0$. 

As in the $G_2$ example, ``integrating in'' the M5-brane state 
requires the introduction of a bare Chern-Simons coupling 
$k=-\ft12$, in order to cancel the induced Chern-Simons coupling 
when it is subsequently integrated out.
The bosonic 
part of the low-energy effective action is therefore given by,
\be
{\cal L}_{\Sfive}=\int_{\mr^{1,2}}\ \frac{1}{e^2}F\wedge^\star F+
\frac{k}{4\pi}A\wedge F + |{\cal D}q|^2 +\phi^2|q|^2
\label{lagsfive}\ee
Notice that the D-term contribution to the potential energy, 
given in equations \eqn{dterm1} and \eqn{dterm2} for previous 
examples, is absent in this case. This is because, in 
three dimensional theories with ${\cal N}=1$ supersymmetry, 
D-terms arise from scalar, rather than vector, multiplets and, 
in the present case, the $\phi$ field is non-normalizable. 
In contrast, note that, even in the 
$e^2\rightarrow\infty$ limit, the gauge field retains a single 
derivative kinetic term. 

Although the Lagrangian \eqn{lagsfive} is not parity invariant 
due to the presence of the Chern-Simons term, after integrating out 
the fermionic superpartner of $q$, this term is canceled and 
parity is restored at low energies as required by the 
discussion in Section 2.6. 

What happens as the ${\bf S}^5$ shrinks to zero size? 
With only ${\cal N}=1$ supersymmetry for 
protection, it is difficult to make any concrete statements about 
the strong coupling physics. Nevertheless, we shall present a 
consistent picture which passes several tests. We conjecture that, 
as suggested classically, the state arising from the M5-brane 
wrapping ${\bf S}^5$ becomes light in this limit. From Section 2, 
we have learnt that the manifold $X=\Sfive$ can undergo a geometrical 
transition to the topologically distinct manifold 
$\mq\cong{\mathbb R}^4\times {\mathbb C}{\bf P}^2$. 
From the perspective of the low-energy dynamics, the natural 
interpretation of this is as a transition onto the Higgs branch 
\cite{Strominger,gms} through condensation of M5-branes,
\be
|q|^4 \sim {\rm Vol}({\mathbb C}{\bf P}^2)
\label{koreascore}\ee
Since the state which condenses is non-BPS, it is hard to 
prove explicitly that this occurs. Still, there 
are several checks we can perform to see if such an interpretation 
holds water. Firstly, 
consider the symmetries of M-theory compactified on
$X\cong{\mathbb R}^4\times {\mathbb C}{\bf P}^2$. The relevant 
cohomology groups are:
\be
H^2(Y;U(1))=U(1)_J\quad,\quad H^2(X;U(1))=U(1)_J
\label{arsenalvschelseatoday}\ee
implying that a global $U(1)_J$ symmetry is left unbroken on this
branch. This is indeed the case on the Higgs branch of our low-energy 
effective theory, if we identify $U(1)_J$ with the action on the 
dual photon. 

Further agreement arises from examining the various extended 
objects that exist in M-theory on $\mq$, arising from wrapped M5 or
M2-branes. At first sight, it appears that we may wrap an
M5-brane over the four-cycle ${\mcp^2}\subset \mq$ to get a domain
wall in $d=3$. However, there is
a subtle obstruction to doing this. Specifically, on an M5-brane
worldvolume $W$ propagates a chiral two-form with self-dual field strength
$T$. This satisfies the relation \cite{wittenm5}
\be
dT= G\mid_W
\ee
In particular, if there is a cohomologically non-trivial $G$-flux over
$W$, one cannot wrap a five-brane. But this is precisely the case for
$X=\mq$ since the membrane anomaly requires a half-integral flux of
$G$ over $\mcp^2$. Thus there is no wrapped five-brane.

However, we are free to wrap an M2-brane over ${\mcp^1}\subset
\mcp^2$. This non-BPS state has a semi-classical mass proportional 
to the volume of $\mcp^1$, and is electrically charged under the global 
$U(1)_J$ symmetry with the identification 
\eqn{arsenalvschelseatoday}. From the D6-brane perspective of Section 
2, this state is a fundamental string stretched between the two 
disjoint D6-branes depicted in Figure (\ref{hlfigb}C). What does this 
state correspond to from the perspective of our low-energy gauge theory? 
The fact that it is charged under $U(1)_J$ implies that it must be 
a vortex. Vortices in non-supersymmetric Maxwell-Chern-Simons 
theories have been studied in the literature \cite{khare}. The 
mass of such an object is expected to be proportional to $|q|^2$. 
Comparing to the mass of the M2-brane state, we are led to the 
relationship \eqn{koreascore}.

\begin{figure}

\setlength{\unitlength}{0.9em}
\begin{center}
\begin{picture}(22,7)

\qbezier(0,6)(5,6)(10,4)
\qbezier(0,2)(5,2)(10,4)
\qbezier(0,6)(1,4)(0,2)
\qbezier(0,6)(-1,4)(0,2)
\put(-7.5,3.5){$X=\Sfive$}

\put(10,4){\line(5,2){6}}
\put(10,4){\line(5,-2){6}}
\put(17,6){$\mq=\mr^3\times\mcp^2$}
\put(18.5,4.7){$k=+\ft12$}
\put(17,1){$\mq=\mr^3\times\mcp^2$}
\put(18.5,-0.3){$k=-\ft12$}
\put(28,3.5){Parity}
\qbezier(26,6.3)(29,3.65)(26,1.3)
\put(26,6.3){\line(1,0){0.5}}
\put(26,6.3){\line(0,-1){0.5}}
\put(26,1.3){\line(1,0){0.5}}
\put(26,1.3){\line(0,1){0.5}}

\end{picture}\end{center}
\caption{The moduli space of M-theory on the $Spin(7)$ 
cone over the Aloff-Wallach space $N_{1,-1}=SU(3)/U(1)$. 
The G-flux is measured by $\int_{\mcp^2}G/2\pi=k$. Parity 
acts as a reflection in the horizontal axis.}
\label{figend}
\end{figure}
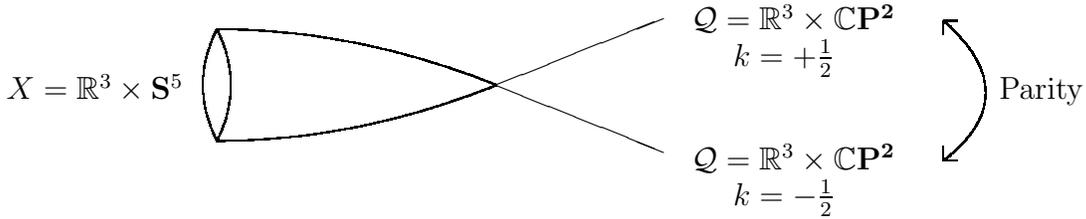

We turn now to the question of parity breaking. In the 
previous section, we have argued that the transition onto the manifold 
$\mq\cong\mr^3\times\mcp^2$ spontaneously breaks parity due to the 
presence of non-zero G-flux. Since we have chosen the parity 
transformation \eqn{parity} to act on $\mr^{1,2}$, this should 
be reflected in parity breaking of the low-energy theory.
This occurs naturally in our picture due to the Chern-Simons 
term coupling in \eqn{lagsfive}. Recall that on the Coulomb branch, 
this was canceled upon integrating out the Dirac fermionic 
superpartner of $q$ --- call it $\psi$ --- to result in a parity 
invariant theory. However, if the mass of $\psi$ vanishes, or 
indeed becomes negative, this cancellation no longer occurs and 
parity is broken. 

While we have presented a plausible scenario for the low-energy 
description of the topology changing transition, we should point out 
that, with such little supersymmetry, other possibilities 
exist. For example, the $U(1)$ gauge symmetry may, instead,  
be broken by a Cooper pair of $\psi$. Parity is then, once again, 
broken in the Higgs branch as required. This is somewhat similar 
to the phenomenon of p-wave superconductivity.

Finally, we may sketch the moduli space of M-theory on the $Spin(7)$ 
cone over the Aloff-Wallach space $N_{1,-1}$. It consists of 
three branches: a two-dimensional Coulomb branch corresponding to 
the geometry $X\cong\Sfive$, and two, one-dimensional Higgs 
branches, related by parity, each corresponding to the 
geometry $\mq\cong\mr^4\times\mcp^2$. The resulting picture is drawn in 
Figure (\ref{figend}).

\section{Geometric Quotients}

In this section we describe the proposed $Spin(7)$ conifold
transition in M-theory in terms of two equivalent, but rather
different, type IIA duals. These dual pictures correspond to
choosing different M-theory circles on which to reduce.

In general, given a $U(1)$ isometry of $X$, one
can choose to embed the M-theory circle along the $U(1)$ orbits.
If $U(1)$ acts freely --- that is,
there are no fixed points of the circle action --- then the quotient
space $X/U(1)$ is a
manifold, and M-theory on $X$ is dual to type IIA string theory on the
quotient $X/U(1)$ with, in general, a
non-trivial RR 1-form potential $A^{RR}$ and dilaton field $\varphi$. The
field strength $F^{RR}=dA^{RR}$ is then
interpreted geometrically as the curvature of the M-theory circle bundle
\be
U(1)\hookrightarrow X \rightarrow X/U(1)\label{fibration}
\ee
A reduction to type IIA also exists when the $U(1)$ action
has a fixed point set $W$ of codimension four in $X$. Then the
quotient $X/U(1)$ may be
given a manifold structure, with $W$ embedded as a codimension three
submanifold. The circle fibration (\ref{fibration}) now degenerates over $W$,
which is interpreted as the locus of a D6-brane in type IIA string
theory. The field strength $F^{RR}/2\pi$ is no longer closed. Rather, its
integral over a small two-sphere ${\bf S}^2$
linking $W$ in $X/U(1)$ is equal to one.

We shall find two interesting type IIA duals of our conifold transition in
M-theory, corresponding to different choices of M-theory circle, which we refer to as the 
B-picture and L-picture \cite{aw,GS,gt}. Let us
briefly summarize these dual pictures. We begin with the B-picture.

The starting point is to consider type IIA string theory on the
asymptotically conical $G_2$ manifold which is an $\mr^3$ bundle over $\mcp^2$
\be
\mr^3 \times \mcp^2
\ee
The low energy effective theory has $\mathcal{N}=2$
supersymmetry in $d=3$, and was discussed in section 3.1.2. One may now 
wrap\footnote{One must also
include a suitable half-integral flux of the gauge field strength on the
D6-brane, as discussed earlier.} a space-filling D6-brane around the
calibrated bolt $\mcp^2$, thus breaking $\mathcal{N}=2$ to $\mathcal{N}=1$, and 
lift the whole configuration to M-theory. In
M-theory, this configuration is described by the $Spin(7)$ manifold $\mq\cong
\mr^4\times \mcp^2$. 

The conifold transition in M-theory corresponds, in the B-picture, 
to shrinking the
$\mcp^2$ bolt to zero size, and blowing up a different copy of
$\mcp^2$. Thus, in the B-picture, the conifold transition looks like
a flop transition in which one copy of $\mcp^2$ collapses and another
blows up. The space-filling D6-brane turns into a single unit of RR
flux through $\mcp^1\subset\mcp^2$. This brane/flux transition is illustrated in Figure 3.

The second type IIA dual, which we refer to as the L-picture, was
discussed in section 2. Here the M-theory conifold transition
corresponds, in type IIA string theory, to a transition of coassociative
submanifolds in \emph{flat space},
with D6-branes wrapped on the calibrated submanifolds. The
coassociative submanifolds in question were first constructed in the
seminal paper of Harvey and Lawson \cite{HL}. In order
to produce this dual picture, we simply choose a different M-theory
circle on which to reduce. 

In the remainder of the section we describe in detail how these two
pictures emerge. In section 4.1 we describe the B-picture, and its
relation to the quaternionic projective plane. In section 4.2 we
construct the quotients $X/U(1)\cong \mr^7$ explicitly. The topology
of $X$ is completely encoded in the fixed point set $L$, as
exemplified by equations (\ref{homo}). We
also discuss how the symmetries of M-theory on $X$ reduce to
symmetries of the L-picture.


\subsection{Brane/Flux Transitions}

In this subsection we will describe more precisely the duality
between the proposed $Spin(7)$ conifold transition and the B-picture
dual outlined above. Roughly speaking, in the B-picture, the conifold 
transition corresponds to a transition in which D6-branes are replaced
with RR 2-form flux. This type of transition is by now familiar. However,
there are some additional subtleties in our case. As we shall see
presently, there is a curious relation between all of the $Spin(7)$
manifolds discussed in this paper and the quaternionic projective
plane. We therefore begin with a discussion of $\mhp^2$.

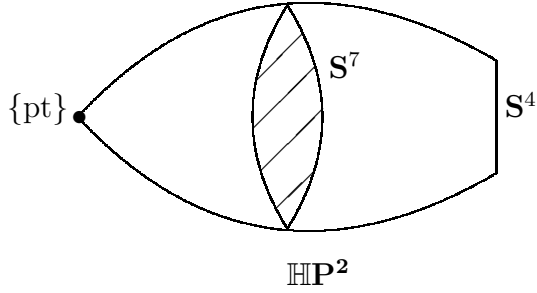
\begin{figure}
\setlength{\unitlength}{0.9em}
\begin{center}
\begin{picture}(25,11)

\qbezier(5,5)(11.5,12)(20,7)
\qbezier(5,5)(11.5,-2)(20,3)
\put(20,3){\line(0,1){4}}

\put(4.8,4.7){$\bullet$}
\put(2.5,5){$\{\mathrm{pt}\}$}
\put(20.3,5){${\bf S}^4$}
\put(12.5,-1){$\mhp^2$}

\put(14,6.4){${\bf S}^7$}

\qbezier(12.5,9)(10,5)(12.5,1)
\qbezier(12.5,9)(15,5)(12.5,1)
\put(11.4,6.3){\line(1,1){1.7}}
\put(11.3,4.6){\line(1,1){2.2}}
\put(11.6,3.1){\line(1,1){2.2}}
\put(12.1,1.7){\line(1,1){1.4}}

\end{picture}\end{center}
\caption{A foliation of $\mhp^2$ space by ${\bf S}^7$ principal
orbits. The two special orbits consist of a point and a 4-sphere.}
\label{figsp}
\end{figure}

\begin{figure}
\setlength{\unitlength}{0.9em}
\begin{center}
\begin{picture}(25,11)

\qbezier(5,7)(12.5,11)(20,7)
\qbezier(5,3)(12.5,-1)(20,3)
\put(5,3){\line(0,1){4}}
\put(20,3){\line(0,1){4}}

\put(3.5,5){${\bf S}^5$}
\put(20.3,5){$\mcp^2$}
\put(12.5,-1){$\mhp^2$}

\put(14,6.4){$U(3)$}
\put(14,6){\line(1,0){2.4}}
\put(14,5){$U(1)^2$}

\qbezier(12.5,9)(10,5)(12.5,1)
\qbezier(12.5,9)(15,5)(12.5,1)
\put(11.4,6.3){\line(1,1){1.7}}
\put(11.3,4.6){\line(1,1){2.2}}
\put(11.6,3.1){\line(1,1){2.2}}
\put(12.1,1.7){\line(1,1){1.4}}

\end{picture}\end{center}
\caption{A foliation of $\mhp^2$ space by $U(3)/U(1)^2$ principal
orbits. The two special orbits are and ${\bf S}^5$ and $\mcp^2$.}
\label{fighp}
\end{figure}
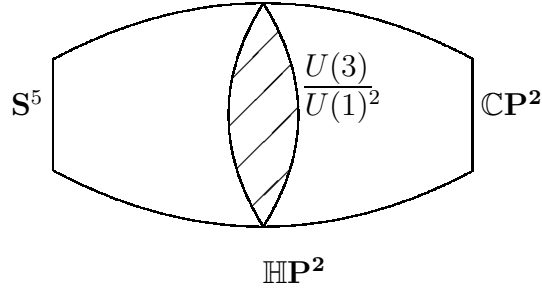

Our starting point is to consider the orbit structure of the
quaternionic projective plane
${\mhp^2} = Sp(3)/Sp(2)\times Sp(1)$ under the two subgroups
$Sp(2)\times Sp(1)$ and $U(3)$ of Sp(3). Notice that these are also the
isometry groups of the $Spin(7)$ manifolds in table \ref{table}. In both
cases the generic orbit of the action on $\mhp^2$ is codimension one. On rather
general grounds, one therefore knows that there will be two special orbits of
higher, and generally unequal, codimension. The generic orbit is
then necessarily a sphere bundle over each of the special orbits. Filling
in each sphere bundle and gluing back-to-back gives a construction of
the manifold compatible with the group action. The orbit structures in
each case is illustrated in Figures \ref{figsp} and \ref{fighp}.

Consider the orbit structure of
$\mhp^2$ under the subgroup $Sp(2)\times
Sp(1)\subset Sp(3)$. This is illustrated in Figure \ref{figsp}. If we denote 
homogeneous coordinates on $\mhp^2$ as
$(u_1,u_2,u_3)$ with $u_i\in\mh$, then the two special orbits
are the point $(0,0,1)$ and the copy of $\mhp^1={\bf S}^4$ consisting
of the points $(u_1,u_2,0)$. The generic orbit ${\bf S}^7$ is the
distance sphere\footnote{In fact, the weak $G_2$ metric
on the squashed seven-sphere, up to homothety, is embedded in the
quaternionic projective plane in this way, for appropriate geodesic
distance \cite{DNP}.} from the point $(0,0,1)$. 

Using this information, we may now construct the B-picture for the $Spin(7)$ manifold 
$\Sigma^-{\bf S}^4$. 
The latter may be obtained from $\mhp^2$ by simply deleting the
special orbit $(0,0,1)$. The
isometry group of the $Spin(7)$ manifold $\Sigma^-{\bf S}^4$ is
precisely $Sp(2)\times Sp(1)$. If we now take the $U(1)$
subgroup given by $U(1)_c\equiv U(1)\subset Sp(1)$ in the last factor of
$Sp(2)\times Sp(1)$, then the fixed point set is the special orbit
${\bf S}^4$. The quotient space is therefore given by\footnote{In the
remainder of the paper we choose orientation conventions such that the
$G_2$ manifold is the bundle of anti-self-dual two-forms over $B$.}
\be
\Sigma^-{\bf S}^4/U(1)_c \cong \Lambda^-{\bf S}^4
\ee
Since the ${\bf S}^4$ descended from a fixed point set, in type IIA we
have a D6-brane wrapped on the bolt of the $G_2$ manifold
$\Lambda^-{\bf S}^4$, the bundle of anti-self-dual 2-forms over 
${\bf S}^4$. The low energy physics of type IIA string theory on this $G_2$
manifold (without the D6-brane) was discussed in section 3.1.2.

Consider now the orbit structure of $\mhp^2$ under $U(3)\subset
Sp(3)$. This is illustrated in Figure \ref{fighp}. In this case, discussed in
\cite{aw}, the generic orbit is the Aloff-Wallach space
$N_{1,-1}=U(3)/U(1)^2$, where $U(1)^2$ is generated by elements of the
form $\mathrm{diag}(\mu,\lambda,\lambda^{-1})\in U(3)$. The two
special orbits are ${\mcp^2} = U(3)/U(2)\times U(1)$
and ${\bf S}^5=U(3)/U(1)\times SU(2)$.
The copy of $\mcp^2$ is merely the subset
of $\mhp^2$ in which all the homogeneous coordinates are purely
complex. If we delete this special orbit from $\mhp^2$,
we obtain a manifold which
is an $\mathbb{R}^3$ bundle over the ${\bf S}^5$ special orbit. This
is in fact the underlying manifold for the $Spin(7)$ geometry of interest.
On the other hand,
deleting the ${\bf S}^5$ special orbit from $\mhp^2$ gives the
manifold $\mq$, which is an $\mathbb{R}^4$ bundle over the $\mcp^2$
special orbit.

To get the B-picture, we consider dividing out by the circle action on $\mhp^2$
generated by the diagonal subgroup
$U(1)_D\subset U(3)$. The fixed point set of this circle action is precisely the special orbit
$\mcp^2$. The generic orbit descends to a copy of the so-called twistor space of
$\mcp^2$, which is the coset-space $U(3)/U(1)^3$. This is also the sphere bundle of
$\Lambda^-\mcp^2$. 

If we delete the fixed copy of $\mcp^2$, we obtain
a \emph{free} action of $U(1)=U(1)_D$ on $\mathbb{R}^3\times {\bf
S}^5$. The
$U(1)$ acts on the ${\bf S}^5$ special orbit by a Hopf map over a
\emph{dual} copy of $\mcp^2$, given by
$\widetilde{\mcp^2}=U(3)/U(1)\times U(2)$. The quotient is therefore
\be
(\Sfive)/U(1)_D \cong \Lambda^-\mcp^2\label{s5quotient}
\ee
with a single unit of RR two-form flux through $\mcp^1\subset
\mcp^2$. Deleting the special orbit ${\bf S}^5$
from $\mhp^2$ and taking the quotient also gives
\be
\mq/U(1)_D \cong \Lambda^-\mcp^2\label{cp2quotient}
\ee
but now the fixed point set $\mcp^2$ becomes a D6-brane wrapped on
the zero-section of $\Lambda^-\mcp^2$. We thus have a picture of the
transition in which D6-branes are replaced with RR flux. Notice that
in this asymptotically conical case, the dilaton blows up at infinity,
so that the type IIA solution is not really valid at large distance. 
However, there exist asymptotically locally conical versions of the two
metrics for which the dilaton stabilizes to a finite value at infinity.
An ALC $Spin(7)$ metric on $\mq$ was constructed in \cite{GS}.

Notice that although the right hand sides of
(\ref{s5quotient}) and (\ref{cp2quotient}) are diffeomorphic, they are
not the ``same'' manifold. The bolt of (\ref{cp2quotient}) came from
the fixed $\mcp^2$, whereas the bolt of (\ref{s5quotient}) came from a
dual copy $\widetilde{\mcp^2}$. These are not the same copy of
$\mcp^2$. Put another way, given a $G_2$ cone on the twistor space
$SU(3)/U(1)^2$, there are three choices of $\mcp^2$ that we may make to
form the resolution $\Lambda^-\mcp^2$, permuted by the Weyl group
$\Sigma_3$ of $SU(3)$. If we start with a D6-brane wrapped on the $\mcp^2$
bolt of the $G_2$ manifold $\Lambda^-\mcp^2$, then as we shrink the bolt
to zero size, another copy of $\mcp^2$ blows up (namely $\widetilde{\mcp^2}$), with the D6-brane
replaced with RR-flux through $\mcp^1 \subset \widetilde{\mcp^2}$.


\subsection{D6-branes on Coassociative Submanifolds of $\mathbb{R}^7$}

In this subsection we describe the construction of the L-picture
discussed in section 2. The aim is to identify the appropriate $U(1)$
subgroup such that $X/U(1)\cong \mr^7$, along with the corresponding fixed point
sets, and also to determine how the symmetries in M-theory are
realized in the L-picture.

Since the construction of these quotients is a little involved, it is
useful at this stage to give a brief summary of the approach that we
will take. The initial problem is to find the appropriate $U(1)$
isometry along which to embed the M-theory circle. In fact, this
is related in a curious way to the B-picture, as described at the end
of this introduction. The
quotient itself is constructed in much the same way as the quotients
in \cite{aw}. Roughly speaking, one foliates the manifold $X$ by a
family of $U(1)$-invariant submanifolds. This family is acted on by a certain $Sp(1)$ subgroup of
the symmetry group of $X$. After we take the quotient by $U(1)$ to
reduce to type IIA on $\mr^7$, this $Sp(1)$ action describes the
sweeping out of $\mathbb{R}^7$ in a form of generalized polar
coordinates. This is precisely the Harvey and Lawson action
(\ref{HLaction}). Recall that this 
$Sp(1)$ is a subgroup of $G_2$ which
preserves the decomposition $\mr^7 = \mathrm{Im} \mh
\oplus \mh$. If one takes an appropriate curve in the
$r-s $ plane, where $s$ and $r$ are radial coordinates on each factor
in the decomposition of $\mathbb{R}^7 = \mathrm{Im} \mathbb{H}
\oplus \mathbb{H} $, then under the action of
$Sp(1)$ we sweep out a coassociative 4-fold $L$, as described in
section 2. At the same time, this $Sp(1)$ action sweeps out the fixed
point set $L$ of the circle action. In mathematical terms, we have
therefore constructed an $Sp(1)$-equivariant map from $X$ to
$\mr^7$. On $X$ this $Sp(1)$ is simply part of the symmetry group. On
$\mr^7$ it's the Harvey and Lawson action (\ref{HLaction}).

\subsubsection*{A Mysterious Duality}

Before we proceed, we pause to point out a curious relation between the L and B-pictures which
emerges via the embedding in $\mhp^2$ described in the last
subsection. Firstly, it is a bizarre
enough fact
that the explicitly known AC $Spin(7)$ and $G_2$ manifolds, together
with their isometry groups, are
related to $\mhp^2$ at all. We have already discussed the $Spin(7)$ case
in this
paper, and two of the three AC $G_2$ manifolds and their relation to $\mhp^2$ was discussed in
\cite{aw}. However, in the $Spin(7)$ case, we also have the following
curious fact. The $U(1)$ subgroup of $Sp(3)$ that produces the
B-picture for the $Spin(7)$ manifold $\Sigma^-{\bf S}^4$ is the \emph{same}
$U(1)$ subgroup that produces the L-picture for the $Spin(7)$ manifold
$\mq$, namely $U(1)=U(1)_c$. Moreover, the converse is also true! That is, the
$U(1)$ subgroup $U(1)_D$ that produces the B-picture for the $Spin(7)$
manifold $\mq$ also produces the L-picture for the $Spin(7)$ manifold
$\Sigma^-{\bf S}^4$. This is most peculiar, and it is not clear
to us whether or not there is any deep underlying reason for this fact.

\subsubsection{The Cone over  $SO(5)/SO(3)$}

The asymptotically conical $Spin(7)$ manifold $X=\Sigma^-{\bf S}^4$ was
first constructed in \cite{gary}, and is a resolution of the cone on
the weak $G_2$ holonomy squashed seven-sphere $Y$. There is a single
modulus $a>0$ corresponding to the radius of the ${\bf S}^4$ bolt. The isometry group is
$Sp(2)\times Sp(1)$. A generic principal orbit $Y={\bf S}^7$  may therefore be viewed as the coset space

\be
Y = Sp(2) \times Sp(1)_c / Sp(1)_{a+c}\times Sp(1)_b
\ee

where we have labelled $Sp(1)_a \times Sp(1)_b \subset Sp(2)$ and $Sp(1)_{a+c}$ denotes
the diagonal subgroup of $Sp(1)_a\times Sp(1)_c$. The seven-sphere $Y$ then fibers over
${\bf S}^4 = Sp(2) \times Sp(1)_c / Sp(1)_a \times Sp(1)_b \times
Sp(1)_c$ with fibers being copies of ${\bf S}^3 = Sp(1)_a \times Sp(1)_b \times
Sp(1)_c / Sp(1)_{a+c}\times Sp(1)_b$. This is the quaternionic Hopf fibration.

In this notation, the $U(1)$ subgroup we require to produce the
quotient (\ref{quotient}) is given by the diagonal
$U(1)_D=U(1)_{a+b+c}\subset Sp(1)_{a+b+c}$. 

The circle action on $\mh^3\setminus\{0\}$ is free, but when we
descend to $\mhp^2$ the special orbit $\mcp^2$ under the action of
$U(3)$ is fixed. The circle action on the ${\bf S}^5$
special orbit merely rotates around the Hopf fibers over the dual copy
of $\mcp^2$. This was described in the last subsection in relation to the
B-picture for the other $Spin(7)$ geometry. 

The quotient of $\mhp^2$ by this
action was shown to be ${\mhp^2}/U(1)_D ={\bf S}^7$ in
\cite{aw}. The manifold $X$ is obtained by
deleting the point $(0,0,1)$ from $\mhp^2$. This descends
to a point in ${\bf S}^7$, and we have therefore shown that
\be\Sigma^-{\bf S}^4/U(1)_D = {\bf S}^7\setminus \{\mathrm{pt}\} =
\mathbb{R}^7
\ee
The fixed point set is given by deleting the point $(0,0,1)$ from the
fixed special orbit $\mcp^2$ to give
\be
L = {\mcp^2}\setminus \{\mathrm{pt}\} = H^1\label{s4fps}
\ee
where $H^1$ denotes the total space of the spin bundle of ${\bf
S}^2$. By supersymmetry, this embedding will be coassociative with
respect to some $G_2$ structure on $\mr^7$.

The symmetry group of M-theory on $X$ consists entirely of geometric
symmetries $Sp(2)\times Sp(1)$. There are no symmetries associated with
the $C$-field in this case since $H^2(X;U(1))$ is trivial. When we
pass to type IIA, we shall find that the symmetry group gets broken to
\be
Sp(2)\times Sp(1) \mapsto U(1)_J\times Sp(1) \times U(1)\label{s4symmetry}
\ee
The $U(1)_J$ is associated with the M-theory circle, and is of course
just $U(1)_D$ in this case. We would like to understand the action of
the remaining factor of $Sp(1)\times U(1)\sim U(2)$ on the type IIA
geometry. Specifically, this will be a symmetry of the embedding of $L=H^1$ in
$\mr^7$. In order to understand this, and its relation to the
Harvey and Lawson geometries, it will be convenient to
understand the quotient just constructed in a rather different way.

The idea is to foliate $X$ by a two-parameter family of invariant
$6$-manifolds whose quotients by $U(1)_D$ may be identified with a
parameterization of $\mr^7$ in terms of generalized polar
coordinates. Specifically, if one denotes
\be
\mathbb{R}^7 = \mathbb{R}^3 \oplus \mathbb{R}^4 = \mathrm{Im}
\mathbb{H} \oplus \mathbb{H}
\ee
then for a unit imaginary quaternion $x\in \mathrm{Im}\mathbb{H}$, we define
\be
\mathbb{R}^5(x) = (tx,y) \quad \mathrm{where\ } t\in\mathbb{R}, y \in
\mathbb{H}\label{polar1}
\ee
As the unit vector $x$ varies over the unit two-sphere, the
spaces $\mathbb{R}^5(x)$ sweep out $\mathbb{R}^7$. Notice that $\pm x$
give the same five-space.

More precisely, this process of sweeping out $\mathbb{R}^7 =
\mathrm{Im}\mathbb{H}\oplus \mathbb{H}$
is achieved by fixing some value of $x$ and then acting with $Sp(1)$
via the Harvey and Lawson action (\ref{HLaction}). This action on $\mathbb{R}^7=\mathrm{Im}\mathbb{O}$
preserves the $G_2$ structure and corresponding splitting into
$\mathrm{Im} \mathbb{H} \oplus \mathbb{H}$. Notice that our fixed
point set (\ref{s4fps}) is topologically the same as that of the Harvey and
Lawson submanifold (\ref{h1}). The deformation parameter $\rho$, which
measures the size of the ${\bf S}^2$ bolt, corresponds to the
deformation parameter $a$ of the $Spin(7)$ geometry which is essentially
the radius of the ${\bf S}^4$ bolt. As we take both parameters to zero, we obtain
a conical geometry in each case: a cone on the weak $G_2$ squashed
seven-sphere in M-theory and a coassociative cone on a squashed
three-sphere for the D6-brane worldvolume in type IIA.

It remains to identify the $Sp(1)\subset G_2$ action of Harvey and
Lawson with part of the unbroken symmetry group in
(\ref{s4symmetry}), and also to understand the action of the
additional $U(1)$ factor. To do this we will follow a
similar route to \cite{aw}, although we will not spell out all of the
details.

Let us view the ${\bf S}^4$ bolt of $X$ as the unit sphere in
$\mathbb{R}^5$. Then $Sp(2)\subset Sp(2)\times Sp(1)$ acts on this
four-sphere via the usual action of $Spin(5)\cong
Sp(2)$. Consider the subgroup $U(2)\subset Sp(2)$. Then $U(1)_D$
restricts to the diagonal $U(1)=U(1)_{a+b}$ and we get a decomposition
\be
\mathbb{R}^5 = \mathbb{R}^2 \oplus \mathbb{R}^3\label{r5}
\ee
where the two-plane $\mathbb{R}^2$ is rotated and the copy of
$\mathbb{R}^3$ is fixed. The set
of fixed-points on ${\bf S}^4$ is therefore a copy of ${\bf
S}^2\subset \mathbb{R}^3$, which is also the subset of ${\bf
\mathbb{H}P}^1 = {\bf S}^4$ in which the homogeneous coordinates are complex. When we take the quotient, since ${\bf S}^2$
has codimension two in ${\bf S}^4$, this fixed point set becomes a
boundary in the reduced space. Specifically, ${\bf S}^4/U(1)_D \cong D^3$, the
closed three-disc. The
boundary $\partial D^3$ of this three-disc is thus identified with the fixed
${\bf S}^2$. Conversely, the $Sp(1)\sim SO(3)$ subgroup of
$U(2)\sim U(1)\times SO(3)$ acts canonically on the factor of
$\mathbb{R}^3$ in (\ref{r5}), and thus acts on the fixed two-sphere
${\bf S}^2\subset \mathbb{R}^3$. If we now view this two-sphere
as the unit sphere in $\mathrm{Im}\mathbb{H}$, then $Sp(1)$ acts by
conjugation of quaternions. This is precisely the action we were looking for. Thus
the $Sp(1)$ factor in the unbroken symmetry group (\ref{s4symmetry})
is identified with $Sp(1)\cong SU(2)\subset U(2)$. The remaining
factor of $U(1)$ is identified with $U(1)_c\subset Sp(1)_c$ which acts
trivially on ${\bf S}^4$. We shall discuss this factor further below.

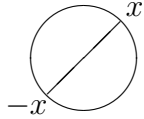
\begin{figure}
\setlength{\unitlength}{0.9em}
\begin{center}
\begin{picture}(25,11)

\put(13,3){\circle{6}}
\put(13,3){\line(1,1){1.3}}
\put(13,3){\line(-1,-1){1.3}}
\put(14.5,4.5){$x$}
\put(10.2,1){$-x$}

\end{picture}\end{center}
\caption{The quotient of ${\bf S}^4$ by $U(1)$ is a 3-disc $D^3$. The
boundary is identified with the fixed ${\bf S}^2$. The image of the
2-sphere ${\bf S}^2_x$ is the line segment joining $x$ to $-x$. }
\label{circle}
\end{figure}

Now fix a point $x\in {\bf S}^2\subset \mathbb{R}^3 = \mathrm{Im}\mathbb{H}$ and consider the
two-sphere ${\bf S}^2_x$ in ${\bf S}^4$ which lives in the three-space
\be
\mathbb{R}^2\oplus \mathbb{R}_x \subset \mathbb{R}^2\oplus\mathbb{R}^3
= \mathbb{R}^5
\ee
Here $\mathbb{R}_x$ denotes the line in $\mathbb{R}^3$ which contains the
point $x\in {\bf S}^2$ and the origin. Then, as we vary $x$ by acting
with $Sp(1)$, we sweep out ${\bf S}^4$ with a two-parameter family of
two-spheres ${\bf S}^2_x$. Notice that for fixed $x$, the image of ${\bf S}^2_x$ in the three-disc $D^3$ is
the line segment joining $[-x,x]$.

Denote the restriction of the $\mathbb{R}^4$ bundle $X=\Sigma^-{\bf S}^4$
to ${\bf S}^2_x$ as $X_x$. As we vary $x$, these invariant 6-manifolds sweep out
$X$. We find that $X_x/U(1)=\mathbb{R}^5(x)$. Thus, as one
varies $x$ over the set of fixed points one sweeps out $\mathbb{R}^7$
with a two-parameter family of five-spaces $\mathbb{R}^5(x)$. In fact,
one can show using similar techniques to \cite{aw} that, as a bundle with $U(1)$ action, $X_x$ is given by
\be
X_x = H^1(1)\oplus H^{-1}(1)
\ee
and that the quotient of this six-manifold by $U(1)$ is indeed
$\mathbb{R}^5(x)$. The details are left as an exercise for the
interested reader, or, alternatively, may be found in \cite{thesis}.

We are now ready to identify the remaining group actions. The zero
section of $X = \Sigma^-{\bf S}^4$ descends to the closed three-disc,
$D^3$. Deleting the fixed point set gives the open three-disc
$\stackrel{\circ}{D^3}$ which is of course diffeomorphic to $\mathbb{R}^3 =
\mathrm{Im}\mathbb{H}$. Thus the quotient of $X$, minus its fixed
points, is
\be
\mathrm{Im}\mathbb{H}\oplus \mathbb{H}\label{open}
\ee
with the second factor coming from the fibers of $X$. The splitting (\ref{open}) is
naturally the spin bundle of
$\mathbb{R}^3=\mathrm{Im}\mathbb{H}$. Indeed, if one deletes the fixed ${\bf
S}^2$ from ${\bf S}^4$, then over the open disc
$\stackrel{\circ}{D^3}$ the chiral spin bundle is simply the spin bundle of
the disc. The $Sp(1)$ action by
conjugation on $\mathrm{Im}\mathbb{H}$ that we have already found
therefore induces an $Sp(1)$ action on the $\mathbb{H}$ factor. This
is the right action \cite{mclean}
\be
y\mapsto y\bar{q}
\ee
Thus the total action of $Sp(1)\subset U(2)$ on $\mathbb{R}^7$ is precisely the
$Sp(1)$ action of Harvey and Lawson (\ref{HLaction}). Including the
fixed points also gives $\mathbb{R}^7$ with (\ref{open}) embedded as a
dense open subset.

Finally, there is the factor $U(1)=U(1)_c$ which acts purely on the
fibers of $X$. The Harvey and Lawson
geometry is $Sp(1)$-invariant by construction, but there is indeed
another $U(1)$ symmetry which corresponds to $U(1)_c$ under our isomorphism. Namely, the action
\be
(x,y)\mapsto (x,\lambda y)\ee
This $U(1)$ is a subgroup of $G_2$ which also preserves the corresponding
decomposition of $\mathbb{R}^7$. It acts by rotating the fibers of $H^1=L$. This is indeed a symmetry of this geometry. A principal
$Sp(1)={\bf S}^3$ orbit is a squashed three-sphere. The isometry group
is $U(2)\sim Sp(1)\times U(1)$, with the $Sp(1)$ acting on the base
${\bf S}^2$ of
the Hopf fibration, and $U(1)$ acting on the fibers. In sum, we have
an action of $Sp(1)\times U(1)$ given by
\be
(x,y)\mapsto (qx\bar{q},\lambda y\bar{q})\ee
Notice that $(-1,-1)\in Sp(1)\times U(1)$ acts trivially, and so the
group that acts effectively here is $(Sp(1)\times U(1))/\mathbb{Z}_2 \cong
U(2)$. The symmetry group (\ref{s4symmetry}) of $X$ is actually more
precisely $(Sp(2)\times Sp(1))/\mathbb{Z}_2$ with the $\mathbb{Z}_2$
generated by $(-1,-1)$. Thus the symmetry breaking may be written more
precisely as
\be
(Sp(2)\times Sp(1))/\mathbb{Z}_2 \mapsto U(1)_J \times U(2)\ee
This completes our analysis.


\subsubsection{The Cone over $SU(3)/U(1)$}

The asymptotically conical $Spin(7)$ manifolds $\mq$ and
$\Sfive$ are both resolutions of the cone on the
weak $G_2$ holonomy Aloff-Wallach space, $N_{1,-1}=SU(3)/U(1)$. Numerical
evidence for the existence of these solutions was given in \cite{garycohom}. The isometry group
is $U(3)$. In each case, there is a single modulus $a>0$
corresponding to the size of the ${\bf \mathbb{C}P}^2$ or ${\bf S}^5$
bolt, respectively.

Our aim in this section is to find a $U(1)$ subgroup of $U(3)$ such that the
quotient spaces may be identified with $\mathbb{R}^7$ (\ref{quotient}). We shall find
that the fixed point sets in each case are given, respectively, by
\be
L = H^1\cup \mathbb{R}^4\ \ \ \ \ \ \ {\rm and}\ \ \ \ \ \ 
L = {\bf S}^3 \times \mathbb{R}
\ee
In the conical limit $a\rightarrow 0$, we therefore have
\be
L\rightarrow C({\bf S}^3\cup {\bf S}^3) = \mathbb{R}^4\cup
\mathbb{R}^4
\ee
Remarkably, these are precisely the coassociative submanifolds
(\ref{niceone}) and (\ref{lfirstres}) considered in section (2.2). If
we place a D6-brane on the coassociative submanifold (\ref{niceone})
and lift to M-theory, we obtain the $Spin(7)$
manifold $\mq$. The symmetry group associated with the C-field in
M-theory is given by $H^2(\mq;U(1))\cong U(1)$ which becomes the axial
$U(1)$ on the D6-branes (the diagonal $U(1)$ decouples as usual). The
circle reduction breaks the geometric symmetry group to
\be
U(3)\mapsto U(2)\times U(1)_J\label{cp2symmetry}
\ee
If we write $U(2)\times U(1)\subset U(3)$ in the obvious way, we shall
find that $U(1)_J$ is given by the $U(1)$ factor, and the $U(2)$ symmetry
of the Harvey and Lawson geometry gets identified with the $U(2)$
factor. Notice this is the same $U(2)$ subgroup of $Sp(3)$ found in
the last section, but now we have $U(1)_J = U(1)_c$. The $Sp(1)$ that sweeps out $L$ is given by
$Sp(1)\cong SU(2)\subset U(2)$. Notice also that the coassociative plane
$\mathbb{R}^4=\mathbb{H}$ is just the copy of $\mathbb{H}$ in
$\mathrm{Im}\mathbb{H}\oplus \mathbb{H}$ at $s=0$, and this is also
swept out under
\be
(x,y)\mapsto (q x \bar{q}, y\bar{q})
\ee
on taking $x=0$ (rather than $x=\epsilon$ a unit vector, which describes the
$H^1$ component of $L$).

Conversely, if we place a D6-brane on (\ref{lfirstres})
and lift to M-theory, we obtain the $Spin(7)$
manifold $\Sfive$. The symmetry group
$H^2(X;U(1))$ associated with the C-field in
M-theory is now trivial, which corresponds to the fact that there is
only one connected component of $L$ in this resolution.
The circle reduction again breaks the geometric symmetry group
as in (\ref{cp2symmetry}).

In the remainder of the paper, we give an explicit construction of the
L-picture quotients.

\emph{\bf The First Resolution}

We would like to construct the isomorphism $\mq/U(1)_c\cong
\mathbb{R}^7$ where the codimension four fixed point set is
$L=H^1\cup\mathbb{R}^4$, thus making contact with the coassociative
geometry described in section 2. The bundle $\mq$ is a chiral spin
bundle (or, more precisely, a $\mathrm{spin}^c$ bundle), but in this subsection we shall find the
following description of $X=\mq$ more useful.

The universal quotient bundle $\mq$ fits into the following short
exact sequence of vector bundles
\be
0\rightarrow L^{(2)}_{\mathbb{C}} \rightarrow {\bf
\mathbb{C}P}^2\times \mathbb{C}^3 \rightarrow \mq \rightarrow 0
\ee
Here $L^{(2)}_{\mathbb{C}}$ denotes the canonical complex line bundle
over ${\bf \mathbb{C}P}^2$. It is a sub-bundle of the trivial bundle ${\bf
\mathbb{C}P}^2\times \mathbb{C}^3$ defined by
\be
L^{(2)}_{\mathbb{C}} = \{(l,z)\in {\bf
\mathbb{C}P}^2\times \mathbb{C}^3 \mid z\in l\}
\ee
That is, the fiber of $L^{(2)}_{\mathbb{C}}$ above the point $l\in
{\bf \mathbb{C}P}^2$ is the
complex line through $l$. In fact, one may make a similar definition
of $L^{(n)}_{\mathbb{K}}$ as the canonical line bundle over ${\bf
\mathbb{K}P}^n$, for any of the associative normed division algebras $\mathbb{K}=\mathbb{R},\mathbb{C},\mathbb{H}$. For
example, the topology of $X$ in the last subsection is $L^{(1)}_{\mathbb{H}}$.

The group $U(3)$ acts on ${\bf \mathbb{C}P}^2\times \mathbb{C}^3$ by
the diagonal action on each factor, and, in this way, $U(3)$ acts on
the line bundle $L^{(2)}_{\mathbb{C}}$ and therefore on the quotient $\mq$, which
we may take to be $(L^{(2)}_{\mathbb{C}})^{\bot}$ with respect to a
flat metric on $\mathbb{C}^3$.

The $U(1)$ subgroup we require is $U(1)=U(1)_c$ generated by elements
$(1,1,\lambda^2)$ inside the maximal torus of $U(3)$. The fixed point
set on the zero section ${\bf \mathbb{C}P}^2$ consists of two
components. In terms of homogeneous
coordinates on ${\bf \mathbb{C}P}^2$, the circle action fixes the point
$A=(0,0,1)$, together with a copy of $B={\bf \mathbb{C}P}^1={\bf S}^2$
given by the points $(z_1,z_2,0)$.

The fiber of $L^{(2)}_{\mathbb{C}}$ above the fixed point $A$ is
$(0,0,z)\in\mathbb{C}^3$ and is acted on by our $U(1)$ subgroup with
weight 2. Conversely, the $\mathbb{R}^4=\mathbb{C}^2$ fiber of $\mq$
above $A$ consists of the points $(z_1,z_2,0)\in \mathbb{C}^3$, which
is clearly fixed under the $U(1)$ action. Thus the total fixed point set above
$A$ is a copy of $\mathbb{R}^4$.

The total fixed point set above $B={\bf \mathbb{C}P}^1$ in ${\bf \mathbb{C}P}^2\times
\mathbb{C}^3$ is ${\bf \mathbb{C}P}^1\times \mathbb{C}^2$. Over $B$,
the line bundle $L^{(2)}_{\mathbb{C}}$ restricts to
$L^{(1)}_{\mathbb{C}}$, and we have the following short exact sequence
\be
0\rightarrow L^{(1)}_{\mathbb{C}} \rightarrow {\bf
\mathbb{C}P}^1\times \mathbb{C}^2 \rightarrow E \rightarrow 0
\ee
The quotient bundle $E$ is a line bundle and is just $(L^{(1)}_{\mathbb{C}})^{-1}$,
which is thus the total fixed point set in $\mq$ above $B$. This line
bundle has first Chern class one, and so is also $H^1$. In sum, the
total fixed point set is
\be
L = \mathbb{R}^4 \cup H^1
\ee
Since $L$ has codimension four in $X=\mq$, the quotient will be a
manifold, and is in fact diffeomorphic to $\mathbb{R}^7$. Rather as
before, we shall find that
$\mq$ may be foliated by a two-parameter family of invariant
$6$-manifolds whose quotients by $U(1)$ may be identified with a
parameterization of $\mathbb{R}^7$ in terms of generalized polar
coordinates. Specifically, if one denotes
\be
\mathbb{R}^7 = \mathrm{Im}\mathbb{H} \oplus \mathbb{H}
\ee
then for a unit vector $x\in \mathrm{Im}\mathbb{H}$, we define
\be
\mathbb{R}^5_+(x) = (tx,y) \quad \mathrm{where\ } t\in\mathbb{R}_+, y \in
\mathbb{H}\label{polar}
\ee
As the unit vector $x$ varies under the natural action of $Sp(1)$, the
half-spaces $\mathbb{R}^5_+(x)$ sweep out $\mathbb{R}^7$, with the
boundary of the half space as axis. In fact, this axis will turn out
to be the fixed $\mathbb{R}^4$ above the point $A$.

The quotient of ${\bf \mathbb{C}P}^2$ by $U(1)$ is again the closed three-disc $D^3$. The image of the point $A$ lies at the center
of $D^3$, and the boundary $\partial D^3$ of $D^3$ is identified with
the fixed ${\bf \mathbb{C}P}^1$: $\partial D^3 \cong B$. For each
fixed point $x \in B$, we now define the two-sphere ${\bf S}^2_x$ to be the copy of ${\bf
\mathbb{C}P}^1\subset {\bf \mathbb{C}P}^2$ containing the points $x$
and $A$. The two-sphere $B$ is acted on by the subgroup $Sp(1)\cong
SU(2)\subset U(2)$ in (\ref{cp2symmetry}). In this way we
sweep out ${\bf \mathbb{C}P}^2$ with a two-parameter family of
two-spheres, with axis $A$. If we view $B$ as the unit sphere in
$\mathrm{Im}\mathbb{H}$, then this $Sp(1)$ action is by conjugation of
quaternions. Notice that the diagonal $U(1)=U(1)_{a+b}$
acts trivially on $B$. This $U(1)$ will become
the extra $U(1)$ symmetry in the Harvey and Lawson geometry.

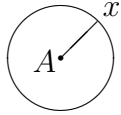
\begin{figure}
\setlength{\unitlength}{0.9em}
\begin{center}
\begin{picture}(25,11)
\put(13,3){\circle{20}}
\put(13,3){\line(1,1){1.3}}
\put(13,3){\circle*{0.2}}
\put(14.5,4.5){$x$}
\put(12,2.5){$A$}

\end{picture}\end{center}
\caption{The quotient of $\mcp^2$ by $U(1)$ is also a 3-disc $D^3$. The
boundary is identified with the fixed ${\mcp^1}={\bf S}^2$, and the
center is the fixed point $A$. The image of the
2-sphere ${\bf S}^2_x$ is the line segment joining $A$ to $x$. }
\label{circle2}
\end{figure}

We again define $X_x$ to be the restriction of
$X = \mq$ to ${\bf S}^2_x$, so that the corresponding $6$-manifolds
sweep out $\mq$, with axis $\mathbb{R}^4_A = X_A$, as $x$
varies over the fixed ${\bf S}^2 = B$. The quotient
spaces $X_x/U(1)$ will turn out to be half-spaces $\mathbb{R}^5_+(x)$
which sweep out $\mathbb{R}^7$ as $x$ varies.

Let us fix $x \in B$. Without loss of generality, we may take
$x=(1,0,0)\in B\subset {\bf \mathbb{C}P}^2$. Each $6$-manifold $X_x$
is a $\mathbb{C}^2$ bundle over ${\bf S}^2_x$. This will decompose
into the sum of two line bundles. Thus, as a bundle with $U(1)$
action, we must have
\be
X_x = H^k_x(m)\oplus H^l_x(n)
\ee
for some integers $k,l,m,n$. In order to work out this splitting,
we may look at the total weights of the $U(1)$ action over the two fixed
points $x$ and $A$.

The fiber of $X_x$ above the north pole $x$ is given by points in
$\mathbb{C}\oplus\mathbb{C}=\{(0,z_2,z_3)\}$. The first copy is fixed,
and the second is acted on with weight 2. Thus the weights on the
tangent space and the two copies of $\mathbb{C}$ above the north pole
$x$ are respectively
\be
(2,0,2)
\ee
The fiber above the south pole $A$ is fixed, so the weights are
\be
(-2,0,0)
\ee
This completely determines the integers $k,l,m,n$ to be $0,1,0,1$, 
respectively. We therefore find that a trivial direction splits off to give
\be
X_x = H^1_x(1)\times \mathbb{C}_x
\ee
where the trivial factor $\mathbb{C}_x$ is fixed under $U(1)$ and the
$H^1_x$ factor is rotated. This splitting is also consistent with the fact
that the first Chern class of $X_x$ should be the generator of
$H^2({\bf S}^2_x;\mathbb{Z})\cong\mathbb{Z}$. Notice that the
fixed $\mathbb{C}_x$ above the point $x \in B$ is the fiber of the
fixed $H^1_B$
above the point $x$. However, above the point $A$ (which is contained
in every ${\bf S}^2_x$) this $\mathbb{C}_x$ lies in the fixed
$\mathbb{R}^4=X_A$. Conversely, the rotated fiber of $H^1_x$ above $x$
coincides with the rotated fiber $\mathbb{C}_B$ above $x$.

The quotient space is given by
\be
X_x/U(1) = (H^1_x(1)/U(1)) \times \mathbb{C}_x= \mathbb{R}^3_+(x)
\times \mathbb{C}_x = \mathbb{R}^5_+(x)
\ee
where we have used the fact that $H^1(1)/U(1)=\mathbb{R}^3_+$
\cite{aw}. The boundary of the half-space $\mathbb{R}^5_+(x)$ is the fixed copy of $\mathbb{H}$,
which is $\{0\}\times \mh$ in  $\mathbb{R}^7=\mathrm{Im}\mathbb{H}\oplus \mathbb{H}$. The diagonal
$U(1)$ subgroup of $U(2)$ acts on the fixed $H^1$ by rotating the
fiber with weight one (it acts trivially on the zero section), and
acts on the fixed $\mathbb{R}^4=\mathbb{C}\oplus \mathbb{C}$ with
weights $(1,1)$. This is the same action that we found in the last
subsection, and thus we identify
this $U(1)$ with the extra $U(1)$ symmetry of the Harvey and Lawson
geometry.

\emph{\bf The Second Resolution}

To complete the picture, we would like to construct the isomorphism
$\Sfive/U(1)\cong \mathbb{R}^7$ with codimension four fixed
point set $L={\bf S}^3\times \mathbb{R}$.

The isometry group is
again $U(3)$ with the M-theory circle being $U(1)=U(1)_c$. The bolt ${\bf S}^5$ is acted on by the $U(3)$ symmetry group in the
obvious way, viewing ${\bf S}^5$ as the unit sphere in
$\mathbb{C}^3$. The $U(1)$ action then decomposes
\be
\mathbb{C}^3 = \mathbb{C}^2\oplus \mathbb{C}
\ee
with the copy of $\mathbb{C}$ rotated and the $\mathbb{C}^2$
fixed. Hence the fixed point set on ${\bf S}^5$ is a copy of ${\bf
S}^3$, the unit sphere in $\mathbb{C}^2$. The restriction of the
$\mathbb{R}^3$ bundle $\Sfive$ to ${\bf S}^3$ is isomorphic to the
product space ${\bf S}^3 \times \mathbb{R}^3$. Our $U(1)$ acts on this
space. Specifically, above each fixed point on the three-sphere, we
get a copy of $\mathbb{R}^3$ on which $U(1)$ acts. There are
essentially only two choices. Either the whole fiber is fixed, or else
we get a splitting
\be
\mathbb{R}^3 = \mathbb{C}\oplus \mathbb{R}
\ee
with the $\mathbb{C}$ rotated (with some weight) and the factor of
$\mathbb{R}$ fixed. Since we know from the
previous subsection that the fixed point set on the
sphere bundle of $\Sfive$ is ${\bf S}^3\cup{\bf S}^3$, this rules out
the first possibility, and we conclude that the total fixed point set
is
\be
L = {\bf S}^3\times \mathbb{R}
\ee
in agreement with the Harvey and Lawson geometry. 
Now fix a point $w\in {\bf S}^3$ and define the two-sphere ${\bf
S}^2_w$ as the unit two-sphere in
\be
\mathbb{R}_w\oplus\mathbb{C}\subset\mathbb{C}^2\oplus\mathbb{C}=\mathbb{C}^3
\ee
where $\mathbb{R}_w$ is the line in $\mathbb{C}^2=\mathbb{R}^4$
through the points $\pm w$. The restriction of the $\mathbb{R}^3$
bundle $X=\Sfive$ to ${\bf S}^2_w$ must split as
\be
X_w = H^k_w(n)\times \mathbb{R}(w)
\ee
for some bundle with $U(1)$ action over ${\bf S}^2_w$ given by
$H^k_w(n)$. We won't actually need to determine this bundle
precisely. We simply observe that, since $L$ has codimension four, the
quotient is a manifold, and the only\footnote{The case $(k,n)=(1,\pm 1)$
is also ruled out since this would require that the entire $\mathbb{R}^3$
fiber above either the north or south pole of ${\bf S}^2_w$ be fixed, which we know
is not the case.} values of $(k,n)$ for which
this is possible are given by $(k,n)=(2,0)$ or $(0,2)$. In either case the
quotient is $H^k(n)/U(1)=\mathbb{R}^3$, and we conclude that
\be
X_w/U(1) = \mathbb{R}^3(w)\times \mathbb{R}(w) = \mathbb{R}^4(w)
\ee
As $w$ varies over the fixed ${\bf S}^3$, these four-spaces sweep out
$\mathbb{R}^7$ in generalized polar coordinates
\be
\mathbb{R}^7 = \mathbb{R}^4\oplus \mathbb{R}^3
\ee
where for a unit vector $w\in\mathbb{R}^4$ we define
\be
\mathbb{R}^4(w)=(tw,v) \quad\mathrm{where}\ t\in\mathbb{R},\
v\in\mathbb{R}^3
\ee
The symmetry group again breaks according to
\be
U(3)\mapsto U(2)\times U(1)_J
\ee
We may again identify the $U(2)$ action with the Harvey and Lawson
symmetry group. Specifically, we see from the construction above that $U(2)$ acts on the ${\bf S}^3$ factor
via its embedding in $\mathbb{C}^2$. The diagonal $U(1)_D\subset U(2)$
Hopf fibers the three-sphere over a copy of ${\bf S}^2$, and the
$Sp(1)\cong SU(2)\subset U(2)$ part acts transitively on ${\bf
S}^3\cong Sp(1)$. This is precisely the action of the $U(2)$ symmetry group on
the Harvey and Lawson coassociative geometry.

\medskip

\centerline{\bf Acknowledgments}
\noindent

We wish to thank Roman Jackiw, Neil Lambert,
Igor Polyubin, Ashoke Sen, Andrew Strominger, Jan Troost,
Cumrun Vafa, Ashvin Vishwanath, Eric Zaslow and 
especially Bobby Acharya and Edward Witten for useful discussions.
S.G. and D.T. would also like to thank the Isaac Newton Institute 
for Mathematical Sciences, Cambridge, UK, and S.G. would further 
like to thank the New High Energy Theory Center at 
Rutgers University, for kind hospitality during the course of this work.
This research was conducted during the period S.G.
served as a Clay Mathematics Institute Long-Term Prize Fellow.
The work of S.G. is also supported in part by grant RFBR No. 01-02-17488,
and the Russian President's grant No. 00-15-99296.
D.T. is a Pappalardo fellow and is grateful to the 
Pappalardo family for their kind support. The work of D.T. also 
supported in part by funds provided by the U.S. Department of Energy
(D.O.E.) under cooperative research agreement \#DF-FC02-94ER40818.

\appendix{Geometry of the $Spin(7)$ Conifold}

In this appendix we describe the geometry
of the $Spin(7)$ cone on $SU(3)/U(1)$ and its complete resolutions
\be
\mathbb{R}^4 \times {\bf \mathbb{C}P}^2
\quad, \quad \Sfive
\ee
Even though the existence of a $Spin(7)$ metric in both cases
is suggested by the dual configuration of D6-branes (see Section 2),
the explicit AC metric is not known. The most systematic and convenient
way to find this metric appears to be via the technique developed
by Hitchin \cite{Hitchin}, which was already applied with great success
in the case of $G_2$ manifolds \cite{GYZ, garysix}. Therefore, one particular
goal of the discussion below will be to explain this technique
and compare the results with what is known in the literature.

In general, given a compact 7-manifold $Y$, the problem is
to find a complete $Spin(7)$ metric on a manifold $X$ with
principal orbits being copies of $Y$. To do this, one
picks a family of 4-forms with a fixed homology class on $Y$:
\be
\rho (x_1, \ldots, x_r) \in \Omega^4_{exact} (Y)
\ee
where $x_1, \ldots, x_r$ are parameters (below functions of $t$).
One can choose $x_i$ such that:
\be
\rho = \sum_i x_i u_i
\ee
where each $u_i$ is exact, {\it i.e.} can be written in the form:
\be
u_i = d (v_i), \quad v_i \in \Omega^3 (Y) / \Omega^3_{closed} (Y)
\ee
Then, the following (indefinite) bilinear
form is non-degenerate \cite{Hitchin}:
\be
Q (u_i , u_j) = \int_Y u_i \wedge v_j
\ee

Given $\rho \in \Lambda^4 V^*$, one can define its Hodge dual
$\sigma \in \Lambda^3 V \otimes \Lambda^7 V^*$.
If $\rho$ has the explicit form $\rho_{ijkl} dx^i dx^j dx^k dx^l$,
then we can write $\sigma$ as $\sigma^{ijk}$.
Taking any $v,w \in V^*$, one can construct a top degree
7-tensor:
\be
v_a \sigma^{aij} w_b \sigma^{bkl} \sigma^{mnp} \in (\Lambda^7 V^*)^2
\ee
Or, one can think of it as a map:
\be
H \colon V^* \to V \otimes (\Lambda^7 V^*)^2
\ee
Evaluating $\mathrm{det} H \in (\Lambda^7 V^*)^{12}$,
one can define:
\be
\phi (\rho) = \vert \mathrm{det} H \vert^{1/12}
\ee
By taking the total volume, one can define the following
functional of $x_1 (t), \ldots, x_r (t)$:
\be
V(\rho) = \int_Y \phi (\rho)
\ee

Once we have $Q(u_i, u_j)$ and $V (\rho)$, we can write
down the gradient flow equations:
\be
{dx_i \over dt} = - Q^{(-1) ij} {dV \over dx^j}
\ee
which gives the desired metric with $Spin(7)$ holonomy.
Indeed, according to \cite{Hitchin} solutions to these equations
define the $\mathrm{Spin}(7)$ structure on the 8-manifold $X$:
\be
\Psi = dt \wedge * \rho + \rho
\ee

Now, let's see how this works in the case we are interested in,
namely, when $Y$ is the Aloff-Wallach space
$$
N_{1,-1}\cong SU(3)/U(1)
$$
To describe the geometry of this space more explicitly,
we define left-invariant 1-forms $L_A^{\ B}$ on $SU(3)$ (with $A=1,\ldots,3$)
satisfying $L_A^{\ A}=0$, $(L_A^{\ B})^{\dagger}=L_B^{\ A}$, together
with the exterior algebra
\be
dL_A^{\ B}=iL_A^{\ C}\wedge L_C^{\ B}
\ee
One must now split the generators into those that lie in the coset
$SU(3)/U(1)$ and those that lie in the denominator $U(1)$.
In particular, one must specify the $U(1)$ generator $Q$.
In the case of $Y=N_{1,-1}$ we have
\be
Q \equiv -L_1^{\ 1}+L_2^{\ 2}
\ee
One may now introduce the 1-forms
\be
\sigma\equiv L_1^{\ 3}, \quad \Sigma\equiv L_2^{\ 3}, \quad \nu\equiv
L_1^{\ 2}\label{forms}
\ee
together with the $U(1)$ generator
\be
\lambda \equiv L_1^{\ 1}+L_2^{\ 2}
\ee
Notice that $\lambda$ and $Q$ generate the maximal torus of
$SU(3)$. Finally, since the forms (\ref{forms}) are complex, one may
split them into real and imaginary parts
\be
\sigma\equiv \sigma_1+i\sigma_2, \quad \Sigma\equiv \Sigma_1+i\Sigma_2,
\quad \nu\equiv \nu_1+i\nu_2
\ee
In order to find possible resolutions of the $Spin(7)$ cone on
$Y=N_{1,-1}$, one needs to find a basis of left-invariant
4-forms $u_i$. Since the 4-forms $u_i$ are also required
to be exact, one starts with left-invariants 3-forms $v_i$,
which in the present case are defined by the condition
\be
v([Q, g_1], g_2, g_3) + v(g_1, [Q, g_2], g_3) + v(g_1, g_2, [Q, g_3]) =0
\ee
for all $g_i \in su(3)/u(1)$.
In our case, $Y=N_{1,-1}$, we find six left-invariant
3-forms $v_i$, two of which turn out to be closed.
Therefore, we end up with only four independent
left-invariant exact 4-forms $u_i = d v_i$:
\begin{eqnarray}
u_1 & = & d(\lambda \sigma_1 \sigma_2) \nonumber \\
u_2 & = & d(\lambda \Sigma_1 \Sigma_2) \nonumber \\
u_3 & = & d(\lambda \nu_1 \nu_2) \nonumber \\
u_4 & = & 4 (- \nu_1 \nu_2 \sigma_1 \sigma_2
+ \nu_1 \nu_2 \Sigma_1 \Sigma_2 + \sigma_1 \sigma_2 \Sigma_1 \Sigma_2)
\end{eqnarray}
Using the definition of the bilinear form $Q$ one can easily compute
\be
Q = \pmatrix{
0 & -2 & -2 & 4 \cr
-2 & 0 & 0 & 4 \cr
-2 & 0 & 0 & -4 \cr
4 & 4 & -4 & 0}
\ee
Finally, we define the $U(1)$ invariant 4-form $\rho$ in terms of the basis $u_i$:
\be
\rho = x_1 u_1 + x_2 u_2 + x_3 u_3 + x_4 u_4
\ee
Following the general prescription,
we now have to compute the invariant functional $V(\rho)$
and derive a system of first-order gradient flow equations
that follow from this $V(\rho)$.
However, since in total we have only four independent functions $x_i$,
under a suitable change of variables the resulting system is guaranteed
to be equivalent to the system of differential equations obtained
for the following metric ansatz
\be
ds_8^2 = dt^2 + a(t)^2(\sigma^2_1+\sigma_2^2)+b(t)^2(\Sigma_1^2
+\Sigma_2^2)+c(t)^2(\nu_1^2+\nu_2^2)+f(t)^2\lambda^2
\label{metric}
\ee
The requirement of $Spin(7)$ holonomy requires the functions to satisfy
the following system of first order differential equations,
\begin{eqnarray}
\frac{\dot{a}}{a} & = &
\frac{b^2+c^2-a^2}{abc} \nonumber \\
\frac{\dot{b}}{b} & = &
\frac{a^2+c^2-b^2}{abc}-\frac{f}{b^2}
\nonumber \\
\frac{\dot{c}}{c} & = &
\frac{a^2+b^2-c^2}{abc}+\frac{f}{c^2}
\nonumber \\
\frac{\dot{f}}{f} & = & -\frac{f}{c^2}
+\frac{f}{b^2}
\label{simplesys}
\end{eqnarray}
where $\dot{a}=\frac{da}{dt}$, etc. These were studied by
Cvetic et al. \cite{garycohom}, where a detailed numerical analysis
was performed. They find a family of metrics for
$\mathbb{R}^4 \times {\bf \mathbb{C}P}^2$
which may be expanded near the bolt at $t=0$,
\ba
a&=&t-\ft12(1+q)t^3+\ldots \nn\\
b&=&1+\ft56 t^2+\ldots \nn\\
c&=&1+\ft23 t^2+\ldots \nn\\
f&=&t+qt^3+\ldots
\nn\ea
{}From this we see that $b$ and $c$ are non-vanishing which,
comparing to \eqn{metric}, ensures that a $\mcp^2$
stabilises at the center of the space. Notice that we have fixed the
overall scale of the manifold. The metric functions may
be extended to regular solutions over the whole space
only for the parameter $q$ in the range
$q\geq q_0$ where $q_0$ is a numerical constant. For all values of $q>q_0$, 
the metric is
asymptotically locally conical, with the circle dual to $\lambda$
having finite size at infinity. For the specific value of $q=13/9$,
a complete analytic solution was found in \cite{GS}.
For $q=q_0$ the metric is asymptotically conical.

The second family of asymptotic expansions
around the bolt is given by \cite{garycohom}
\ba
a&=& 1 -\ft13 qt+(1-\ft5{18}q^2)t^2+(\ft7{45}-\ft{167}{810}
q^2)qt^3+\ldots \nn\\
b&=&1+\ft13 qt +(1-\ft{5}{18}q^2)t^2-(\ft{7}{45}-\ft{167}{810}
q^2)qt^3+\ldots \nn\\
c&=& 2t+\ft{4}{27}(q^2-9)t^3+\ldots \nn\\
f&=&q+\ft23 q^3 t^2
\label{sfiveshort}
\ea
In contrast to the previous solution, $a$, $b$ and $f$ are all
non-vanishing at $t=0$, ensuring that the bolt has topology ${\bf S}^5$.
The parameter $q$ now takes values in $0< q\leq q_0\sim 0.87$.
Even though complete $Spin(7)$ metrics of this type are not known
explicitly, our analysis in Sections 2 and 4, and
the numerical analysis in \cite{garycohom}, strongly suggest that
there is a manifold with short-distance asymptotics (\ref{sfiveshort})
and asymptotically locally conical behavior at large distances,
for all values of $q$ apart from $q_0$.
In the latter case, the metric describes an asymptotically
conical manifold with topology $\Sfive$.


\end{document}